\documentclass[aps,preprint,showpacs,amsmath,amssymb,amsfonts,eqnarray,array]{revtex4}
\usepackage{anysize}
\marginsize{1.9cm}{1.9cm}{1.55cm}{1.55cm}
\usepackage{graphicx}
\usepackage{dcolumn}
\usepackage{bm}
\begin{document}


\title{Inelastic electron and light scattering from the elementary electronic excitations in quantum wells:
Zero magnetic field\\}
\author{Manvir S. Kushwaha}
\address
{\centerline {Department of Physics and Astronomy, Rice University, P.O. Box 1892, Houston, TX 77251, USA}}

\date{\today}

\begin{abstract}

The most fundamental approach to an understanding of electronic, optical, and transport phenomena which the
condensed matter physics (of conventional as well as nonconventional systems) offers is generally founded
on two experiments: the inelastic electron scattering and the inelastic light scattering. This work embarks
on providing a systematic framework for the theory of inelastic electron scattering and of inelastic light
scattering from the electronic excitations in GaAs/Ga$_{1-x}$Al$_{x}$As quantum wells. To this end, we start
with the Kubo's correlation function to derive the generalized nonlocal, dynamic dielectric function, and
the inverse dielectric function within the framework of Bohm-Pines' random-phase approximation. This is
followed by a thorough development of the theory of inelastic electron scattering and of inelastic light
scattering. The methodological part is then subjected to the analytical diagnoses which allow us to sense
the subtlety of the analytical results and the importance of their applications. The general analytical
results, which know no bounds regarding, e.g., the subband occupancy, are then specified so as to make them
applicable to practicality. After trying and testing the eigenfunctions, we compute the density of states,
the Fermi energy, the full excitation spectrum made up of intrasubband and intersubband -- single-particle
and collective (plasmon) -- excitations, the loss functions for all the principal geometries envisioned for
the inelastic electron scattering, and the Raman intensity, which provides a measure of the real transitions
induced by the (laser) probe, for the inelastic light scattering. It is found that the dominant contribution
to both the loss peaks and the Raman peaks comes from the collective (plasmon) excitations. As to the
single-particle peaks, the analysis indicates a long-lasting lack of quantitative comparison between theory
and experiments. It is inferred that the inelastic electron scattering can be a potential alternative of the
inelastic light scattering for investigating elementary electronic excitations in quantum wells.


\end{abstract}

\maketitle


\newpage

\section{Introduction}

Scientific advances in any field are generally known to have been advocated on the basis of a sound competition
between the theory and the experiment. The condensed matter physics (CMP), which encompasses curious inquiries
on a vast majority of natural and man-made systems (of solids as well as fluids), is, by no means, an exception
to this notion. This is true for all the conventional (i.e., the three-dimensional) and the non-conventional
(i.e., the low-dimensional) systems being explored within the huge boundaries of CMP, both with and without
subjecting them to the external probes such as electric and/or magnetic fields. The CMP is, in principle, a
study of -ons such as electron, exciton, helicon, magnon, neutron, phonon, photon, plasmon, polaron, roton, ...
etc. These -ons include the ``real particles'' such as electron, the ``quasi particles'' such as plasmon, and the
``dressed particles'' such as polaron. In the language of the quantum mechanics, these -ons are characterized
either as bosons or as fermions -- the former obey the Bose-Einstein statistics whereas the latter the Fermi-Dirac statistics. The experiments performed to observe the response of an externally perturbed system in the CMP involve
either of these -ons such as electron or photon in the respective spectroscopy.

The quantum wells [or, more realistically, a quasi-two-dimensional electron gas (Q-2DEG) for better and broader
range of physical understanding] are fabricated in a manner in which the charge carriers are constrained in one
dimension or are allowed free motion in two dimensions. A Q-2DEG continues to serve as the motherboard for the
still-low-dimensional systems making up the quantum wires (or Q-1DEG) and quantum dots (or Q-0DEG). It may not
be an exaggeration to say that the CMP explored during the past two and a half decades is, by and large, the
physics of the quantum structures of reduced dimensions. And yet, there are numerous fundamental aspects of the
physics that are still missing from the scenario. One of them is the serious and systematic theory of the
inelastic electron scattering from the electronic excitations in the quantum wells (as well as quantum wires, for
that matter). Filling this gap is one of the principal motivations behind this work. To be specific, we pursue a
systematic, self-contained, and thorough development of the theory of inelastic electron scattering (IES) and
inelastic light (or Raman) scattering (ILS) experiments in the quantum wells in the absence of an applied magnetic
field. An exhaustive (historical) account of the electronic, optical, and transport phenomena in the systems of
reduced dimensions can be found in Ref. 1.

The 2D nature of electron gas -- when only the lowest electric subband is occupied (i.e., the electric quantum
limit) -- was first confirmed experimentally by Fowler et al. in 1966 on the Si (100) surface in the presence of
an applied perpendicular magnetic field [2]. The 2D character of the charge densities was first theoretically
explored by Stern who derived 2D analog of Lindhard dielectric function for an arbitrary wave vector and
frequency [3]. Subsequent theoretical development of the study of elementary excitations mostly in the layered
electron gas was slow but steady [4-14]. The early advances on the subject, focusing largely on the inversion
layers, can be seen in a monumental review published by Ando et al. [15]. The following decade saw an extensive
theoretical work on the electronic excitations in the superlattice systems, both compositional and doping, and
both with and without an applied magnetic field [1].

As to the electronic excitations, our main focus here will be on the charge-density excitations in Q-2DEG in a
single quantum well. The generalization of the whole theory to the periodic systems and to include further
effects of an applied electric field, magnetic field, optical phonons, ...etc. is quite straightforward.
The electronic excitations include intrasubband and intersubband single-particle as well as collective (plasmon)
excitations. The two domains (i.e., single-particle and collective) of behavior of a solid state plasma are
separated by a critical length ($\lambda_c$) called Debye length (screening length) in classical (quantum) plasma.
For $\lambda < \lambda_c$ ($\lambda > \lambda_c$) a plasma responds single-particle-like (collectively). The
critical length has another equally important meaning. It is the screening length of the plasma, i.e., the extent
to which an external electrostatic field will penetrate it before being counter-balanced by the induced field due
to polarization of the medium. Every plasma has a characteristic frequency -- the plasma frequency $\omega_p$ --
which sets the scale of its response to time-varying perturbations. The charge-density excitations (CDE) exist
through the direct Coulomb interaction just as the spin-density excitations (SDE) exist through the
exchange-correlation Coulomb interaction. The energy shift of the collective CDE (SDE) from the bare
single-particle transition energy is a direct measure of depolarization and excitonic shift (exchange and
correlation effects) [1].

While ILS experiments have been overwhelmingly used [16-28] to measure the charge density as well as spin-density
excitations in Q-2DEG, the IES experiments have been rarely employed [29-30]. This is somewhat surprising, given
the fact that the theoretical feedback on the IES [31-45] has a comparable long history to that on the ILS [46-60].
Hypothesizing the interaction between the fast charged particle (i.e., the electron) and the metal electrons
describable within the framework of a dielectric approach, Fermi [31] was the first to calculate the stopping
power of matter for fast charged particles. Subsequently, Kramers [32] employed a similar consideration to
calculate specifically the stopping power due to conduction electrons. However, it would be fair to say that both
IES and ILS started receiving considerable attention for the serious practical purposes in the early sixties. The
essential problem with IES [or electron energy-loss spectroscopy (EELS)], or so it looks like, is the thought of
energy resolution concerned with the low-energy excitations in quantum wells that scares. In EELS, the resolution
was significantly less than for the competing optical techniques such as infrared spectroscopy and Raman
spectroscopy. In the latter techniques the resolution is typically about 0.25 meV, whereas in EELS a resolution of
5 meV was considered to be a good result until late eighties. With the increasing complexity of the problems, it
became desirable to have a sensitive method with a better resolution. The technology of spectrometers is now based
on {\em science} and excellent, easy to operate instruments capable of resolution down to 0.3 meV (theoretical
limit) and 0.5 meV (experimentally achieved limit) have been built [61]. In view of this, we believe that
high-resolution EELS (HREELS) could prove to be a potential alternative of the overused optical techniques.

The purpose of the present paper is to develop a comprehensive theory of the inelastic electron and inelastic light scattering in the single quantum wells in the absence of any applied magnetic field. This obviously necessitates a
systematic knowledge of the single-particle and collective (plasmon) excitation spectrum, at least, for the sake of
comparing and justifying the loss peaks in the IES and the intensity peaks in the ILS. To this end, we derive the
required nonlocal, dynamic dielectric function, inverse dielectric function, and other correlation functions in the
framework of Bohm-Pines' full random-phase approximation (RPA) [62]. We ignore the many-body (exchange-correlation)
effects for the sake of simplicity. It is found that the derivation of the probability (or the loss) function for
the IES is defined in terms of the inverse dielectric function, whereas the cross-section (or the Raman intensity)
for the ILS is given in terms of the (reducible) density-density correlation function. The latter requires one to
introduce the double-time retarded Green function whose equation of motion is solved with the rigorous use of the
rules of second quantization, Fourier transforms, and the RPA.

It is observed that the loss features in the IES as well as the intensity peaks in the ILS are caused by the
collective (plasmon) excitations in the quasi-2D electron gas. However, there also exist some (relatively) weak
signals that substantially correspond to the single-particle excitations. These findings are seen to be in full
agreement with the respective experimental observations. This emboldens our confidence in stating that the HREELS
can be a potential alternative of, for instance, Raman spectroscopy.

The rest of the article is organized as follows. In Sec. II, we present the theoretical framework leading to the
derivation of nonlocal, dynamic, dielectric function, screened potential, inverse dielectric function, Dyson
equation, probability function characterizing the inelastic electron scattering, and the cross-section for
inelastic light scattering. We further diagnose some of the results analytically to fully address the solution of
the problem and the related relevant aspects. In Sec. III, we discuss several illustrative examples of, for
example, excitation spectrum comprising of single-particle and collective excitations, the electron energy loss
spectrum, and the Raman intensity. There we also highlight the importance of studying the inverse dielectric
function in relation with the transport phenomena in such quantum systems. Finally, in Sec. IV, we conclude our
finding and suggest some interesting features worth adding to the problem.

\begin{figure}[htbp]
\includegraphics*[width=8cm,height=9cm]{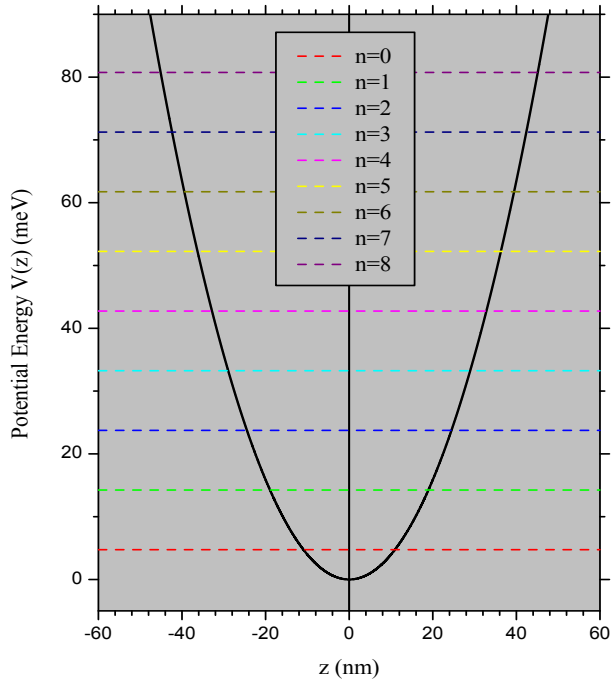}
\caption{(Color online) The 1D parabolic confining potential well: An ideal parabolic potential (such as this)
represents a {\em harmonic oscillator} whose eigenfunction and eigenenergy can be calculated analytically [see,
e.g., Eqs. (3) and (5)]. One feature of a particle that is confined in such a well is that the ground state has
an energy $\epsilon_0=(1/2)\hbar\omega_0$ even at absolute zero of temperature, hence coining the term {\em
zero-point energy}: above this, the equispaced energy steps form a ladder. The eigenfunctions show an
{\em even-odd} alternation just as in the case of a symmetric, square quantum well.}
\label{fig1}
\end{figure}

\section{Methodological Framework}

\subsection{The eigenfunctions and eigenenergies}

We consider a moderate-gap GaAs/Ga$_{1-x}$Al$_{x}$As system with a confining harmonic potential along the z
direction of the conventional 3DEG defined by $V_c(z)=(1/2)m^*\omega^2_0 z^2$ [see, e.g., Fig. 1]. The resulting
system is a typical Q-2DEG with a free electron motion in the x-y plane and the size quantization in the z direction.
For such a system, the single-particle Hamiltonian is expressed as
\begin{equation}
H =\frac{\hat{\boldsymbol p}^2}{2\, m^*} \, + \, \frac{1}{2}\, m^* \,\omega^2_0 \, z^2\, ,
\end{equation}
where $\hat{\boldsymbol p}=-i\,\hbar\nabla_{2D}$ is the momentum operator in the x-y plane. In this situation the
resultant Q-2DEG system can be characterized by the eigenfunction
\begin{equation}
\psi_j({\boldsymbol r})=\frac{1}{\sqrt{A}}\,e^{i\boldsymbol {k}_{\parallel}\cdot\boldsymbol {x}_{\parallel}} \,
\phi_n(z)\, ,
\end{equation}
where ${\boldsymbol r} \equiv ({\boldsymbol x}_{\|}, z)$, ${\boldsymbol x}_{\|}=(x, y)$ is a 2D vector in the direct
space, $A$ the normalization area, $j\equiv {\boldsymbol k_{\|}},n$ the composite index, and the Hermite function
$\phi_n(z)$ is defined as
\begin{equation}
\phi_n(z)=N_n\,e^{-z^2/(2\ell^2_c)}\,H_n(z/\ell_c)\, ,
\end{equation}
where $n$, $N_n=(\sqrt{\pi}\,2^n\,n!\,\ell_c)^{-1/2}$, and $\ell_c=\sqrt{\hbar/(m^*\omega_0)}$ are, respectively,
the subband index due only to the size quantization along the spatial dimension z of the electron gas system, the normalization constant, and the characteristic length of the harmonic oscillator, and the eigenenergy
\begin{equation}
\epsilon_n({k_{\parallel}})= \frac{\hbar^2k^2_{\parallel}}{2m^*} \,+\,\epsilon_n\, ,
\end{equation}
where $k_{\|}=\mid {\boldsymbol k}_{\|}\mid$, ${\boldsymbol k}_{\|}\equiv (k_x, k_y)$ is a 2D wave vector in the
reciprocal space, and $\epsilon_n$ defined as
\begin{equation}
\epsilon_n=(n+\frac{1}{2})\,\hbar\,\omega_0
\end{equation}
is the energy of the $n$th subband. Here $\omega_0$ is the characteristic frequency of the harmonic oscillator.
Note that in the 1D case each energy level corresponds to a unique quantum state and hence the system as such
stands as non-degenerate. It is interesting to add that the great advantage of the harmonic (parabolic) potential
is that one can do a reasonable amount of analytical work to solve, in this case, a 1D Schr\"odinger equation and
deduce the exact form of the wave function [Eq. (3)] and the energy [Eq. (5)] giving one a feel and confidence
about the nature of the calculations involved. However, the general theory for the electronic excitations, IES,
and ILS developed in this paper is independent of any specific model potential confining the charge carriers
along the z direction and making the existence of Q-2DEG feasible.

\subsection{The nonlocal, dynamic dielectric function}

We start with the general expression of the single-particle density-density correlation function (DDCF)
$\chi^0 (...)$ given by [1]
\begin{equation}
\chi^{0} ({\boldsymbol r},{\boldsymbol r'}; \omega)=\sum_{ij}\, \Lambda_{ij}\,\,
\psi^*_i ({\boldsymbol r})\,\psi_j ({\boldsymbol r})\,
\psi^*_j ({\boldsymbol r'})\,\psi_i ({\boldsymbol r'})\, ,
\end{equation}
where the composite index $i,j\equiv k,n$ and symbol $\Lambda_{ij}$ is defined as follows.
\begin{equation}
\Lambda_{ij}= 2\, \frac{f(\epsilon_i)-f(\epsilon_j)}{\epsilon_i-\epsilon_j+\hbar\omega^+} \, ,
\end{equation}
where $f(x)$ is the familiar Fermi distribution function. $\omega^+=\omega+i\gamma$ and small but nonzero
$\gamma$ stands for the adiabatic switching of the Coulomb interactions in the remote past. The factor of
$2$ takes care of the spin degeneracy.

Next, we recall the Kubo's correlation function [1] to write the induced particle density given by
\begin{eqnarray}
n_{in}({\boldsymbol x_{\|}}, z;\omega)
& =& \int d{\boldsymbol x'_{\|}} \int dz'\, \chi^0 ({\boldsymbol x_{\|}}, z; {\boldsymbol x'_{\|}}, z';\omega)\,
V_{tot}({\boldsymbol x'_{\|}}, z';\omega)\nonumber\\
& =& \int d{\boldsymbol x'_{\|}} \int dz'\, \chi ({\boldsymbol x_{\|}}, z; {\boldsymbol x'_{\|}}, z';\omega)\,
V_{ex}({\boldsymbol x'_{\|}}, z';\omega) \, ,
\end{eqnarray}
where $V_{tot}=V_{ex}+V_{in}$ is the total potential, with $V_{ex}$ ($V_{in}$) as the external (induced) potential.
[It should be pointed out that although we use the term potential throughout, we mean it to be the potential energy
unequivocally.] $\chi$ and $\chi^0$ are, respectively, the interacting and the single-particle DDCF and are related
to each other by the Dyson equation [see Fig. 2]
\begin{equation}
\chi ({\boldsymbol r}, {\boldsymbol r'};\omega) = \chi^0({\boldsymbol r}, {\boldsymbol r'};\omega) + \int d{\boldsymbol r^{''}}\,\int d{\boldsymbol r^{'''}}\,
\chi^0({\boldsymbol r}, {\boldsymbol r^{''}};\omega)\, V_{ee}({\boldsymbol r^{''}}, {\boldsymbol r^{'''}})\,
\chi({\boldsymbol r^{'''}}, {\boldsymbol r'};\omega),
\end{equation}
\begin{figure}[htbp]
\includegraphics*[width=8cm,height=4.5cm]{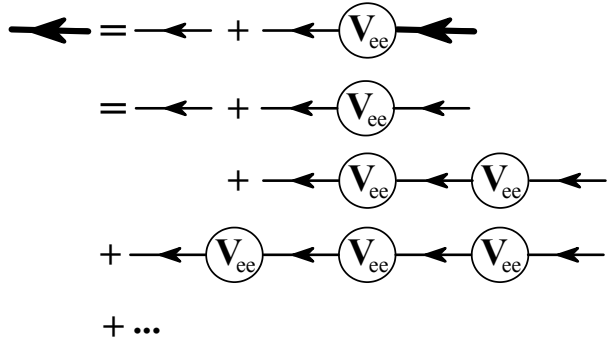}
\caption{The diagrammatic derivation of the Dyson equation: the thick [thin] line represents the reducible [irreducible]
DDCF $\chi (...)$ [$\chi^0 (...)$] in the full RPA. $V_{ee}$ is the binary Coulombic interaction. The arrows indicate
the transition from initial to final spatio-temporal position of the particle.}
\label{fig2}
\end{figure}
where $V_{ee}(...)$ represents the binary Coulomb interactions and is defined as
\begin{equation}
V_{ee}({\boldsymbol r}, {\boldsymbol r'})
=\frac{e^2}{\epsilon_b}\,\frac{1}{\mid {\boldsymbol r}-{\boldsymbol r}'\mid}
=\frac{e^2}{\epsilon_b}\,\frac{1}
{\mid ({\boldsymbol x_{\|}}-{\boldsymbol x'_{\|}})^2+(z-z')^2 \mid^{1/2}}\, ,
\end{equation}
where $-e$ ($e>0$) is the elementary electronic charge and $\epsilon_b$ the background dielectric constant of the
medium hosting the Q-2DEG. Further, the induced potential in terms of the induced particle density is expressed
as
\begin{equation}
V_{in}({\boldsymbol x_{\|}}, z;\omega)=\int d{\boldsymbol x'_{\|}} \int dz'\,
V_{ee}({\boldsymbol x_{\|}}-{\boldsymbol x'_{\|}}; z, z')\,
n_{in}({\boldsymbol x'_{\|}}, z';\omega)
\end{equation}
Equation (11), with the aid of Eqs. (2), (3), (6), (8), and (10), takes the following form.
\begin{eqnarray}
V_{in}({\boldsymbol x_{\|}}, z; \omega)=
&& \frac{1}{A^2}\,\sum_{n n'}\,\sum_{\boldsymbol k_{\|} \boldsymbol k_{\|}'}\,
   \Lambda_{nn'}({\boldsymbol k_{\|}}, {\boldsymbol k_{\|}'}; \omega)\,
  e^{i{\boldsymbol q_{_{\|}}}\cdot{\boldsymbol x'_{\|}}}\,e^{-i{\boldsymbol q_{_{\|}}}\cdot{\boldsymbol x''_{\|}}}\,\nonumber\\
&& \times\,  \int d{\boldsymbol x'_{\|}} \int dz'\int d{\boldsymbol x''_{\|}} \int dz''\,
  V_{ee}({\boldsymbol x_{\|}}-{\boldsymbol x'_{\|}}; z, z')\, \nonumber\\
&& \times\, \phi^*_n (z')\,\phi_{n'} (z')\,\phi^*_{n'} (z'')\,\phi_n (z'')\,
 V_{tot}({\boldsymbol x''_{\|}}, z''; \omega)\, ,
\end{eqnarray}
where ${\boldsymbol k_{\|}'}={\boldsymbol k_{\|}}+{\boldsymbol q_{_{\|}}}$ and ${\boldsymbol q_{_{\|}}}$ is the
momentum transfer. Next, we open the sum over ${\boldsymbol k'_{\|}}$, multiply both sides of Eq. (12) by
$e^{-i{\boldsymbol q'_{_{\|}}}\cdot{\boldsymbol x_{\|}}}$, and integrate with respect to ${\boldsymbol x_{\|}}$.
The result, after replacing the dummy variable ${\boldsymbol q'_{_{\|}}}$ by ${\boldsymbol q_{_{\|}}}$, is
\begin{eqnarray}
V_{in}({\boldsymbol q_{_{\|}}}, \omega; z)=
&& \sum_{n n'}\,
\Pi_{nn'}({\boldsymbol q_{_{\|}}}; \omega)\,\nonumber\\
&& \times\, \int dz'\,\phi^*_{n} (z')\,\phi_{n'} (z')\,V_{ee}(q_{_{\|}}; z, z')\,\nonumber\\
&& \times\, \int dz''\,\phi^*_{n'} (z'')\,\phi_{n} (z'')\,V_{tot}({\boldsymbol q_{_{\|}}}, \omega; z'')\, ,
\end{eqnarray}
where we have made use of the identity
$(2\pi)^2\,\delta ({\boldsymbol k'_{\|}}-{\boldsymbol k_{\|}}-{\boldsymbol q_{_{\|}}})=A \,\delta_{{\boldsymbol k'_{\|}},{\boldsymbol k_{\|}}+
{\boldsymbol q_{_{\|}}}}$. The symbol $\Pi_{nn'}({\boldsymbol q_{\|}}, \omega)$ is defined as
\begin{equation}
\Pi_{nn'}({\boldsymbol q_{\|}}, \omega)
=\frac{1}{A}\, \sum_{\boldsymbol k_{\|}}\, \Lambda_{nn'}({\boldsymbol k_{\|}},{\boldsymbol k'_{\|}}={\boldsymbol k_{\|}}+{\boldsymbol q_{_{\|}}}; \omega)
=\frac{2}{A}\,\sum_{\boldsymbol k_{\|}}\,\frac{f(\epsilon_{{\boldsymbol k_{\|}}n})-f(\epsilon_{{\boldsymbol k'_{\|}}n'})}
               {\epsilon_{{\boldsymbol k_{\|}}n}-\epsilon_{{\boldsymbol k'_{\|}}n'}+\hbar\omega^+}\, .
\end{equation}
In Eq. (13), $V_{ee}(q_{_{\|}}; z, z')$ is the 2D Fourier transform of the binary Coulombic interactions defined
by
\begin{equation}
V_{ee}(q_{_{\|}}; z, z')=\frac{2\pi e^2}{\epsilon_b \, q_{_{\|}}}\,e^{-q_{_{\|}}\mid z-z'\mid}\, .
\end{equation}
Next, let us take the matrix elements of both sides of Eq. (13) between the states $\mid m'>$ and $\mid m>$. The
result is
\begin{equation}
<m' \mid V_{in}(...)\mid m>=\sum_{n n'}\,
\Pi_{nn'}({\boldsymbol q_{\|}}, \omega)\,
F_{nn'mm'}(q_{_{\|}})\, <n'\mid V_{tot}(...)\mid n>\, ,
\end{equation}
where
\begin{equation}
F_{nn'mm'}(q_{_{\|}})=\int dz \int dz' \, \phi^*_{n} (z)\,\phi_{n'} (z)\,
V_{ee}(q_{_{\|}}; z, z')\,\phi^*_{m'} (z')\,\phi_{m} (z')\,
\end{equation}
is the matrix element of the Fourier-transformed Coulombic interactions $V_{ee}(q_{_{\|}}; z, z')$. Let us now
invoke the condition of self-consistency [$V_{tot}=V_{ex}+V_{in}$] on Eq. (16) to write
\begin{equation}
<m' \mid V_{ex}(...)\mid m>=\sum_{n n'}\,[\delta_{nm}\,\delta_{n'm'} -
\Pi_{nn'}(...)\, F_{nn'mm'}(q_{_{\|}})]\, <n'\mid V_{tot}(...)\mid n>\,
\end{equation}
Now, since the external potential and the total potential are correlated through the nonlocal, dynamic dielectric
function $\epsilon (q_{_{\|}},\omega; z, z')$ in the manner
\begin{equation}
V_{ex}({\boldsymbol q_{_{\|}}},\omega; z)=\int dz'\, \epsilon ({q_{_{\|}}},\omega; z, z')\,
V_{tot}({\boldsymbol q_{_{\|}}},\omega; z'),
\end{equation}
we can easily deduce from Eq. (18) that the generalized nonlocal, dynamic dielectric function for the Q-2DEG is
given by
\begin{equation}
\epsilon_{nn'mm'} (q_{_{\|}}, \omega)=\delta_{nm}\,\delta_{n'm'} -
\Pi_{nn'}({\boldsymbol q_{_{\|}}}, \omega)\,F_{nn'mm'}(q_{_{\|}})
\end{equation}
where $\Pi_{nn'}(...)$ and $F_{nn'mm'}(q_{_{\|}})$ are defined, respectively, in Eq. (14) and Eq. (17). The symbol
$\delta_{ij}$ is the usual Kronecker delta defined by $\delta_{ij}=1 (0)$ for $i=j$ $(i\ne j)$. The condition for
the actual instance of the collective excitations is that the self-sustaining plasma oscillations in the electron
density occur. This implies that Eq. (18) has a nonzero solution $V_{in}(...)$ when $V_{ex}(...)=0$. In other
words, the collective excitation spectrum is obtained by the condition of the vanishing of the determinant of the
nonlocal, dynamic dielectric function matrix generated by Eq. (20), i.e., $\mid \epsilon_{nn'mm'}(...)\mid=0$.

\subsection{The inverse dielectric function}

As will be seen later, the cross-section or the loss probability function for the IES (or the EELS) turns out to
be defined in terms of the inverse dielectric function. This requires the derivation of the inverse dielectric
function in a systematic way. A cautionary remark here is that $\epsilon^{-1}(...) \ne 1/\epsilon(...)$ for any
conceptual (or practical) purpose. This is true in spite of the fact that the zeros of the dielectric function
and the poles of the inverse dielectric function must yield exactly identical results. Logically, the key issue
here is that it is a matrix mathematics. Kushwaha and Garcia-Moliner [63] had put forward the process of
systematic derivation of the inverse dielectric function for quasi-n dimensional electron gas [with $n=2, 1, 0$]
systems. We would like to specify the procedure for the Q-2DEG system at hand. To that end, we cast Eq. (13) in
the form
\begin{eqnarray}
V_{ex}({\boldsymbol q_{_{\|}}}, \omega; z) =\!\!\!\int dz'\,[\delta(z-z') &-&
\sum_{n n'}\,
\Pi_{nn'}({\boldsymbol q_{_{\|}}}, \omega)\,\nonumber\\
&&\times\, \int dz''\phi^*_{n} (z'')\,\,V_{ee}(q_{_{\|}}; z, z'')\,\phi_{n'} (z'')\nonumber\\
&&\times\, \phi^*_{n'} (z')\,\phi_{n} (z')]\,V_{tot}({\boldsymbol q_{_{\|}}}, \omega; z').
\end{eqnarray}
Comparing this equation with Eq. (19) yields
\begin{eqnarray}
\epsilon({q_{_{\|}}}, \omega; z, z')=\delta (z-z') &-& \sum_{n n'}\,
\Pi_{nn'}({\boldsymbol q_{_{\|}}}, \omega)\,\nonumber\\
&&\times\, \int dz''\phi^*_{n} (z'')\,\,V_{ee}(q_{_{\|}}; z, z'')\,\phi_{n'} (z'')\nonumber\\
&&\times\, \phi^*_{n'} (z')\,\phi_{n} (z')
\end{eqnarray}
In what follows, we will confine our attention to the process of determining the inverse dielectric function
$\epsilon^{-1}({q_{_{\|}}}, \omega; z, z')$ -- from Eq. (22) -- that satisfies the integral equation
\begin{equation}
\int dz''\, \epsilon^{-1}(z, z'')\,\epsilon (z'', z') = \delta (z-z')
\end{equation}
Let us now define a long range part of the response by
\begin{equation}
L_{nn'}(z)=\int dz'\, \phi_n(z')\,V_{ee}(z,z')\, \phi^*_{n'}(z')\, ,
\end{equation}
and the short range part by
\begin{equation}
S_{nn'}(z')=\phi_n(z')\,\phi^*_{n'}(z')\, ,
\end{equation}
in order to rewrite Eq. (22) in the form [suppressing the ($q_{_{\|}}, \omega$) dependence]
\begin{equation}
\epsilon(z, z')=\delta (z-z') - \sum_{nn'}\,L^*_{nn'}(z)\,\Pi_{nn'}\,S_{nn'}(z')
\end{equation}
Let us transform each pair of subband indices ($nn'$) into a composite index $\mu=\mu_s, \mu_a$, where the subscript
$s (a)$ refers to the symmetric (antisymmetric) function depending on whether $n+n'=$ even or odd. The aforesaid
scheme is quite general and only singles out the symmetric structures from the antisymmetric ones. One can also
choose to use a degenerate Fermi-Dirac statistics as an alternative. There the only non-vanishing elements of the polarizability function $\Pi_{nn'}$ are those of the first $n_i$ rows and $n_i$ columns; where $n_i$ is the number
of occupied subbands. It is noteworthy that this discussion precludes a bit complicated situation where $\phi_n$'s
may become complex, for example. Thus we can cast Eq. (26) in the form
\begin{equation}
\epsilon(z, z')=\delta (z-z') - \sum_{\mu}\,L^*_{\mu}(z)\,\Pi_{\mu}\,S_{\mu}(z')
\end{equation}
where the range of summation is an interval ($a, b$) which can safely be taken to be ($\infty, \infty$) for
generality. We intend to determine $\epsilon^{-1}(z,z')$ given presumably by, say,
\begin{equation}
\epsilon^{-1}(z, z')=\delta (z-z') + \sum_{\nu}\,A^*_{\nu}(z)\,H_{\nu}\,B_{\nu}(z')
\end{equation}
such that the integral Eq. (23) is satisfied. Substituting Eqs. (27) and (28) in Eq. (23) gives
\begin{eqnarray}
\sum_{\mu}\,L^*_{\mu}(z)\,\Pi_{\mu}\,S_{\mu}(z') &-&
\sum_{\nu}\,A^*_{\nu}(z)\,H_{\nu}\,B_{\nu}(z')\nonumber\\
&+& \sum_{\mu\nu}\,A^*_{\nu}(z)\,H_{\nu}\,\Pi_{\mu}\,S_{\mu}(z')\alpha_{\mu\nu}=0
\end{eqnarray}
where
\begin{equation}
\alpha_{\mu\nu}=\int dz\, L^*_{\mu}(z)\,B_{\nu}(z)
\end{equation}
Replace the sum over $\nu$  with $\mu$ -- with no loss of generality -- in the second term on the left-hand side
of Eq. (29) to write
\begin{equation}
\sum_{\mu}\,L^*_{\mu}(z)\,\Pi_{\mu}\,S_{\mu}(z')
+ \sum_{\mu\nu}\,A^*_{\nu}(z)\,H_{\nu}\,
[\Pi_{\mu}\,S_{\mu}(z')\alpha_{\mu\nu} - B_{\nu}(z')\,\delta_{\mu\nu}]=0
\end{equation}
Multiplying this equation with $L^*_{\gamma}(z')$ and integrating over $z'$ gives
\begin{equation}
\sum_{\mu}\,L^*_{\mu}(z)\,\Pi_{\mu}\,\beta_{\gamma\mu}
+ \sum_{\mu\nu}\,A^*_{\nu}(z)\,H_{\nu}\,
[\Pi_{\mu}\,\beta_{\gamma\mu}\alpha_{\mu\nu} - \alpha_{\gamma\nu}\,\delta_{\mu\nu}]=0
\end{equation}
where
\begin{equation}
\beta_{\gamma\mu}=\int dz\, L^*_{\gamma}(z)\,B_{\mu}(z)
\end{equation}
Let
\begin{equation}
B_{\mu}(z)=\lambda_{\mu}\,S_{\mu}(z)\, ,
\end{equation}
so that
\begin{equation}
\alpha_{\mu\nu}=\lambda_{\nu}\,\beta_{\mu\nu}\, ,
\end{equation}
and Eq. (32) takes the form
\begin{equation}
\sum_{\mu}\,[L^*_{\mu}(z)\,\Pi_{\mu}
+ \sum_{\nu}\,A^*_{\nu}(z)\,H_{\nu}\,\lambda_{\nu}\,
(\Pi_{\mu}\,\beta_{\mu\nu} - \delta_{\mu\nu})]\,[\beta_{\gamma\mu}]=0
\end{equation}
Now, it makes sense to argue that either the first or the second factor is zero. Since the second factor
$\beta_{\gamma\mu} \ne 0$, we are left with
\begin{equation}
L^*_{\mu}(z)\,\Pi_{\mu}
= \sum_{\nu}\,A^*_{\nu}(z)\,H_{\nu}\,\lambda_{\nu}\,
(\delta_{\mu\nu}-\Pi_{\mu}\,\beta_{\mu\nu})
\end{equation}
If $\Lambda_{\mu\nu}$ is assumed to be the inverse of $(\delta_{\mu\nu}-\Pi_{\mu}\,\beta_{\mu\nu})$ such that
\begin{equation}
\sum_{\nu}\,(\delta_{\mu\nu}-\Pi_{\mu}\,\beta_{\mu\nu})\,\Lambda_{\nu\gamma}=\delta_{\mu\gamma}\, ,
\end{equation}
then multiplying Eq. (37) by $\Lambda_{\mu\gamma}$ and summing over $\mu$ yields
\begin{eqnarray}
\sum_{\mu}\,L^*_{\mu}(z)\,\Pi_{\mu}\,\Lambda_{\mu\gamma}
&=& \sum_{\nu}\,A^*_{\nu}(z)\,H_{\nu}\,\lambda_{\nu}\delta_{\nu\gamma}\nonumber\\
&=& A^*_{\gamma}(z)\,H_{\gamma}\,\lambda_{\gamma}
\end{eqnarray}
This implies that
\begin{equation}
A^*_{\gamma}(z) = \frac{1}{H_{\gamma}\,\lambda_{\gamma}}\,\sum_{\mu}\,L^*_{\mu}(z)\,\Pi_{\mu}\,\Lambda_{\mu\gamma}
\end{equation}
Equation (28), with the aid of Eqs. (34) and (40), takes the following form.
\begin{equation}
\epsilon^{-1}(z, z')=\delta (z-z') + \sum_{\mu\nu}\,L^*_{\mu}(z)\,\Pi_{\mu}\,\Lambda_{\mu\nu}\,S_{\nu}(z')
\end{equation}
{\em That this is exactly the correct inverse of $\epsilon (z, z')$ can easily be justified by substituting
$\epsilon (z, z')$ from Eq. (27) and $\epsilon^{-1} (z, z')$ from Eq. (41) in Eq. (23)}. Clearly, that is not
all. We are still left with an important question to be addressed: {\em Is $\epsilon^{-1}(z, z')$ in Eq. (41)
the unique inverse of $\epsilon (z, z')$ in Eq. (27)}? In order to make sure, let us answer this question in
negation and suppose that $\kappa^{-1} (z, z')$ [$\ne \epsilon^{-1} (z, z')$] is an another inverse of
$\epsilon (z, z')$. Then there must be a matrix, say, $\Lambda'_{\mu\nu}$ satisfying an identity such as the
one given in Eq. (38), i.e.,
\begin{equation}
\sum_{\nu}\,(\delta_{\mu\nu}-\Pi_{\mu}\,\beta_{\mu\nu})\,\Lambda'_{\nu\gamma}=\delta_{\mu\gamma}\,
\end{equation}
Subtracting Eq. (42) from Eq. (38) piecewise leaves us with
\begin{equation}
\sum_{\nu}\,(\delta_{\mu\nu}-\Pi_{\mu}\,\beta_{\mu\nu})\,(\Lambda_{\nu\gamma}-\Lambda'_{\nu\gamma})=0
\end{equation}
Since the first term is {\em not} zero -- otherwise the identity in Eq. (38) or Eq. (42) makes no sense -- the
second term equated to zero is the proper solution of Eq. (43), i.e. $\Lambda_{\nu\gamma}=\Lambda'_{\nu\gamma}$.
This leads us to inferring that $\kappa^{-1} (z, z') = \epsilon^{-1} (z, z')$. Amen!

\subsection{The screened interaction potential}

A nonlocal, dynamic dielectric function [$\epsilon ({q_{_{\|}}}, \omega; z, z')$] contains the seeds of manifold
fruitful descriptions of physical phenomena. An interesting application of the dielectric formulation of the
many-body theory is to the problem of screening of the electron-electron (e-e) interactions. It is because of the
screening of e-e interactions that the quasi-particle model works as well as it does for the transport phenomena. Qualitatively, each electron in the system behaves like a moving test charge: it acts to polarize its surroundings.
Another electron sees the electron {\em plus} its accompanying time-dependent polarization cloud -- the effective
interaction between the electrons is thus {\em dynamically screened}. In other words, every electron interacts
with other electrons at any distance as though it had a smaller charge: it has been screened by other electrons.
As a result, it is surrounded by a region in which the density of electrons is lower than usual. This region is
typically termed as {\em screening hole}. Viewed from a large distance, this screening hole has the effect of a
coated positive charge which cancels the electric field caused by the ({\em mobile}) electron. The screening
virtually weakens the long-range nature of the Coulombic interactions and acts very strongly to reduce the
correlation effects.

We discuss a full nonlocal and dynamic free-carrier screening effects to be calculated in terms of the screened
interaction potential related to the bare Coulomb potential without any limitation and/or approximation with
respect to the subband structure. For this purpose, we consider two test electrons occupying spatial positions
$\boldsymbol r$ and $\boldsymbol r'$. Their interaction energy is given in terms of the screened Coulomb potential
such that
\begin{equation}
V_s({\boldsymbol r}, {\boldsymbol r}'; t-t')=\int d{\boldsymbol r}''\, \epsilon^{-1}({\boldsymbol r}, {\boldsymbol r}''
; t-t')\, V_{ee}({\boldsymbol r}'', {\boldsymbol r}')\, ,
\end{equation}
where the binary Coulombic interaction term $V_{ee}(...)$ is defined (just as before) by
\begin{equation}
V_{ee}({\boldsymbol r}, {\boldsymbol r}')=\frac{e^2}{\epsilon_b}\,\frac{1}{\mid {\boldsymbol r}-{\boldsymbol r}'\mid}
\end{equation}
Taking the Laplacian [$\nabla^2_{{\boldsymbol r}'}$] of Eq. (41) and using the identity
\begin{equation}
\nabla^2\,V_{ee}({\boldsymbol r}, {\boldsymbol r}')=-4\pi\,e^2\,\delta({\boldsymbol r}-{\boldsymbol r}'), ,
\end{equation}
we obtain, from Eq. (44),
\begin{equation}
\nabla^2\,V_s({\boldsymbol r}, {\boldsymbol r}'; t-t')=-4\pi\,e^2\,\epsilon^{-1}({\boldsymbol r}, {\boldsymbol r}';
t-t')
\end{equation}
This justifies the notion that the determination of the screened Coulomb potential and of the inverse dielectric
function are two {\em equivalent} problems.

Next, we Fourier transform Eq. (44) in the 2D plane and with respect to time to write [suppressing the
(${\boldsymbol q_{_{\|}}}, \omega$) dependence]
\begin{equation}
V_s(z, z')=\int dz''\, \epsilon^{-1}(z, z'')\,V_{ee}(z'', z')
\end{equation}
We also recall the similar Fourier transforms of Eq. (8) and (11) to cast them in the form
\begin{equation}
n_{in}(z)= \int dz'\, \chi^0 (z, z')\,V_{tot}(z')
\end{equation}
and
\begin{equation}
V_{ex}(z)=V_{tot}(z) -\int dz'\, V_{ee}(z, z')\,n_{in}(z')
\end{equation}
where the Fourier transformed
\begin{equation}
\chi^{0}(z, z')=\sum_{nn'}\,\Pi_{nn'}(...)\,\phi^*_{n}(z)\,\phi_{n'}(z)\,\phi^*_{n'}(z')\,\phi_{n}(z')
\end{equation}
where $\Pi_{nn'}(...)$ is just as given in Eq. (14). Equation (50), with the aid of Eq. (49), assumes the form
\begin{eqnarray}
V_{ex}(z)
&=&V_{tot}(z) -\int dz'' \int dz'\, V_{ee}(z, z')\,\chi^0 (z', z'')\,V_{tot}(z'')\nonumber\\
&=&\int dz''[\delta(z-z'')-\int dz'\,V_{ee}(z, z')\,\chi^0{z', z''}]\, V_{tot}(z'')\nonumber\\
&=&\int dz'\, \epsilon(z,z')\,V_{tot}(z')
\end{eqnarray}
where we have redefined the previous $\epsilon(z, z')$ [see Eq. (22)] as follows.
\begin{equation}
\epsilon(z, z')=\delta(z-z') - \int dz''\,V_{ee}(z, z'')\,\chi^0(z'', z')
\end{equation}
Substituting Eq. (53) in Eq. (23) yields
\begin{eqnarray}
\epsilon^{-1}(z, z')
&=&\delta(z-z') + \int dz''\int dz'''\, \epsilon^{-1}(z, z'')\,V_{ee}(z'', z''')\,\chi^0(z''', z')\nonumber\\
&=&\delta(z-z') + \int dz''V_s(z, z'')\,\chi^0(z'', z')
\end{eqnarray}
Equation (48), with the aid of Eq. (54), now assumes, after rearranging the terms, the following form.
\begin{equation}
V_s(z, z')=V_{ee}(z, z') + \int dz'' \int dz'''\, V_{ee}(z,z'')\,\chi^0(z'', z''')\,V_s(z''', z')
\end{equation}
This is the Dyson equation relating the screened interaction potential to the bare Coulombic potential through
the density-density correlation function $\chi^0(z, z')$. Remember, we have, in this section, suppressed the
(${\boldsymbol q_{_{\|}}}, \omega$) dependence of most of the quantities for the sake of brevity. Note that
Eq. (55) can also be derived diagrammatically from Fig. 2 if we identify the thick (thin) line with arrow
referring to the screened (bare Coulomb) potential and encircled $V_{ee}$ is replaced with single-particle
DDCF $\chi^0$.

\subsection{The inelastic electron scattering}

In this section, we focus our attention on the energy loss of a fast charged particle to the plasma medium of a
quasi-2DEG. The theory of EELS -- in the reflection geometry at the surface of a semi-infinite medium -- has
been treated in two different frameworks -- dielectric response theory by Lucas and coworkers [35] and dipole
scattering theory by Persson and coworkers [38] -- with the same basic ingredients. They proceed in two steps:
(i) the incoming fast electron is considered as a classical trajectory, and (ii) the collective excitations are
described in a quantal fashion. More general theories that allow multiple losses or gains and treat the incoming
fast electron as a quantal trajectory have, however, been constructed [36]. In the limit of a weak perturbation
(and small energy losses) these general theories reduce to a simple classical trajectory approach proposed by
Schaich [37]. This limiting approach has two features worthy of attention: first, it is simpler to deal with and
second, it captures the essential physics involved. It may, however, become risky when the energy losses involved
are in the range of several electron volts -- in the conventional solids, for example. Since the quantum structures
(as is the case here) involve energy losses on the order of a few meV, we believe we are just as safe as we ought
to be.

Let us first review some of the basic features that relate to the theory of IES. Since the energy loss is assumed
to be small, the particle is considered to be moving with a uniform velocity such that electron trajectory be
described by
\begin{equation}
{\boldsymbol r}\,(t)={\boldsymbol x}_{\|}(t) + z(t)\,\hat{z}={\boldsymbol v}\,t+{\boldsymbol r}_0
=({\boldsymbol v}_{\|}\,t+{\boldsymbol r}_{0\|}) + (v_z\,t+z_0)\hat{z}\, ,
\end{equation}
where the subscripts on the quantities specify them to be parallel (with subscript $\|$) or perpendicular (with
subscript $z$) to the Q-2DEG. The fast-particle with a charge distribution $\rho({\boldsymbol r}, t)=-e\delta
({\boldsymbol r}-{\boldsymbol r}(t))$ impresses a Coulomb potential
\begin{equation}
V_{ex}({\boldsymbol r}, t)=-e\,\phi_{ex}({\boldsymbol r}, t)=\frac{e^2}{\mid {\boldsymbol r}- {\boldsymbol r}(t)\mid}
\end{equation}
Taking its Laplacian gives
\begin{equation}
\nabla^2\,V_{ex}({\boldsymbol r}, t)=-4\,\pi\,e^2\,\delta({\boldsymbol r}- {\boldsymbol r}(t))\, .
\end{equation}
The problem is addressed in terms of an effective potential $V_{tot}({\boldsymbol r}, t)=V_{ex}({\boldsymbol r}, t)+V_{in}({\boldsymbol r}, t)$, where $V_{in}$ [$V_{ex}$] is given by Eq. (11) [Eq. (57)]. The induced particle density
is defined in terms of an induced potential such as
\begin{equation}
n_{in}({\boldsymbol r}, t)=-\frac{1}{4\,\pi\,e^2}\,\nabla^2[V_{tot}({\boldsymbol r}, t)-V_{ex}({\boldsymbol r}, t)]\, .
\end{equation}
The classical trajectory approach proceeds by noting that the (coherent) incoming electron beam polarizes the
plasma medium of a quasi-2DEG. The induced polarization produces an electric field which exerts a force back on
the electron beam as it approaches the surface of the system. We calculate the total work done by the induced
force to obtain the total energy loss suffered by the electron beam. An appropriate decomposition of the
resulting expression yields the energy distribution of those electrons which suffer an inelastic scattering. We
will not hereinafter use the qualifiers such as coherent, incoming, and beam! It should be made clear that the
low-case (big-case) $v$ refers to the velocity (potential energy). We write the net effect in terms of the energy
loss at the rate defined by
\begin{equation}
\frac{dW}{dt}=-{\boldsymbol v}(t)\cdot {\boldsymbol F}(t)\, ,
\end{equation}
where $t$ is the time, ${\boldsymbol v}$ the velocity, and $F$ the induced force. As $t$ runs from $-\infty$ to
$+\infty$, the incoming electron completes its specular trajectory with its total energy loss given by
\begin{equation}
W=-Re \left [\int^{+\infty}_{-\infty}dt \, {\boldsymbol v}(t)\cdot {\boldsymbol F}(t) \right ]\, .
\end{equation}
If the total energy lost by the particle is cast in the form
\begin{equation}
W=\int d{\boldsymbol q}_{\|} \int d\omega \, \hbar\omega\, P({\boldsymbol q}_{\|}, \omega)\, ,
\end{equation}
the quantity $P({\boldsymbol q}_{_{\|}}, \omega)\,d{\boldsymbol q}_{_{\|}} \,d\omega$ is termed as the probability
that the incoming electron is inelastically scattered into the range of energy losses between $\hbar\omega$ and
$\hbar(\omega+d\omega)$, and into the range of momentum losses parallel to the surface between
$\hbar{\boldsymbol q}_{_{\|}}$
and $\hbar({\boldsymbol q}_{_{\|}}+d{\boldsymbol q}_{_{\|}})$. The angular resolved loss function $P(\omega)$
completely specifies the kinematics of the external electron at the detector.

At the outset, we need to calculate the induced force which is defined by
\begin{equation}
{\boldsymbol F}=\int d{\boldsymbol r}\, n_{in}({\boldsymbol r}, t)\,\nabla V_{tot}({\boldsymbol r}, t)\,.
\end{equation}
This, with the aid of Eq. (59), becomes
\begin{equation}
{\boldsymbol F}=-\frac{1}{4\pi\,e^2}\,\int d{\boldsymbol r}\,
\nabla^2 [V_{tot}({\boldsymbol r}, t)-V_{ex}({\boldsymbol r}, t)]\,\nabla V_{tot}({\boldsymbol r}, t)\,.
\end{equation}
The first term in the integrand representing the so-called self-force is eliminated through the introduction of the
electric field stress tensor. We can understand this in the following way. Since ${\boldsymbol E}=-(1/e)\,\nabla
V_{tot}$ and $\nabla \times {\boldsymbol E}=0$, the first term inside the integrand of Eq. (64) takes the form
\begin{equation}
\nabla^2 V_{tot}\,\nabla V_{tot} \equiv e^2\,[(\nabla.{\boldsymbol E}){\boldsymbol E}-{\boldsymbol E}\times (\nabla\times{\boldsymbol E})]\, .
\end{equation}
After a few algebraic steps, one finds that its x-Cartesian component is given by
\begin{equation}
e^2\,\sum_{\beta}\frac{\partial}{\partial x_{\beta}}\,
(E_{\alpha}\,E_{\beta}-\frac{1}{2}\delta_{\alpha\beta}\,E^2_{\beta})\, ,
\end{equation}
where $\alpha,\beta \equiv x, y, z$; and $\delta_{\alpha\beta}$ is the Kronecker delta. Other two components can be
written analogously. The result is that the tensor divergence of Eq. (66) integrates to zero over all space for a
medium of arbitrary inhomogeneity. Consequently, Eq. (64) reduces to
\begin{equation}
{\boldsymbol F}=\frac{1}{4\pi\,e^2}\,\int d{\boldsymbol r}\,\nabla^2 V_{ex}({\boldsymbol r}, t)\,\nabla V_{tot}({\boldsymbol r}, t)\,.
\end{equation}
Exploiting the translational invariance in the x-y plane of confinement, all quantities (involved in the process)
can be Fourier-transformed with respect to ${\boldsymbol x}_{\|}$ and time, just as before. Equation (67), with the aid of
Eq. (58), assumes the form
\begin{equation}
{\boldsymbol F}=-\nabla V_{tot}({\boldsymbol r}, t) {\large {\mid}_{{\boldsymbol r}={\boldsymbol r}(t)=[{\boldsymbol x}_{\|}(t)+z(t)\hat{z}]=
[({\boldsymbol v}_{\|}\,t+{\boldsymbol x}_{0\|}) + (v_z\,t+z_0)\hat{z}]}}\, .
\end{equation}
The Fourier-transformed solution of Eq. (58) is given by
\begin{equation}
V_{ex}({\boldsymbol q}_{_{\|}},\omega; z)=\frac{2\pi e^2}{q_{_{\|}}}\int dt\,
{\large e^{i[\omega t-{\boldsymbol q}_{_{\|}}\cdot {\boldsymbol x}_{_{\|}}(t)]}\,e^{-q_{_{\|}}\mid z-z(t)\mid}}\, .
\end{equation}
The total potential in the medium outside the quasi-2DEG ($z<0$) is given by
\begin{equation}
V_{tot}({\boldsymbol q}_{_{\|}},\omega; z)=\int dz'\, \epsilon^{-1}({\boldsymbol q}_{_{\|}},\omega; z, z')\,
V_{ex}({\boldsymbol q}_{_{\|}},\omega; z')\, ,
\end{equation}
where
\begin{equation}
\epsilon^{-1}({\boldsymbol q}_{_{\|}},\omega; z, z')=\delta(z-z') + \int dz''\, V_{ee}(q_{_{\|}},z-z'')\,
\chi({\boldsymbol q}_{_{\|}},\omega; z'', z')\, .
\end{equation}
Now, the total potential in the direct space, with the aid of Eqs. (69) and (70), can be written as
\begin{eqnarray}
V_{tot}({\boldsymbol r}, t)
&=&V_{tot}({\boldsymbol x}_{\|}, z, t)=\frac{1}{(2\pi)^3}\int d{\boldsymbol q}_{_{\|}}\,\int d\omega\,
     e^{i({\boldsymbol q}_{_{\|}}\cdot {\boldsymbol x}_{_{\|}}-\omega t)}\, V_{tot}({\boldsymbol q}_{_{\|}}, \omega; z)\nonumber\\
&=& \frac{1}{(2\pi)^3}\int d{\boldsymbol q}_{_{\|}}\,\int d\omega\, e^{i({\boldsymbol q}_{_{\|}}\cdot {\boldsymbol x}_{_{\|}}-\omega t)}\,
\int dz' \epsilon^{-1}({\boldsymbol q}_{_{\|}}, \omega; z, z')\, V_{ex}({\boldsymbol q}_{_{\|}}, \omega; z')\nonumber\\
&=& \frac{e^2}{4\pi^2}\, \int d{\boldsymbol q}_{_{\|}}\, \int d\omega\,
            e^{i({\boldsymbol q}_{_{\|}}\cdot {\boldsymbol x}_{_{\|}}-\omega t)}\,
     \frac{1}{q_{_{\|}}}\,\int dz' \,\epsilon^{-1}({\boldsymbol q}_{_{\|}}, \omega; z, z') \nonumber\\
&&\hspace{4.0cm}\times \,\int dt'\,{\large e^{i[\omega t'-{\boldsymbol q}_{_{\|}}\cdot {\boldsymbol x}_{_{\|}}(t')]}\,
         e^{-q_{_{\|}}\mid z'-z(t')\mid}}
         {{\left\vert\vphantom{\frac{1}{1}}\right.}_{{\boldsymbol r}={\boldsymbol r}(t)}}\, .
\end{eqnarray}
Substituting Eq. (72) in Eq. (68) yields induced force defined by
\begin{eqnarray}
{\boldsymbol F}=-\frac{e^2}{4\pi^2}\,\nabla\, \int d{\boldsymbol q}_{_{\|}}\, \int d\omega\,
            e^{i({\boldsymbol q}_{_{\|}}\cdot {\boldsymbol x}_{_{\|}}-\omega t)}\,\frac{1}{q_{_{\|}}}\,
            \int dz'\, \epsilon^{-1}({\boldsymbol q}_{_{\|}}, \omega; z, z') \nonumber\\
\hspace{4.0cm}\times \,\int dt'\,{\large e^{i[\omega t'-{\boldsymbol q}_{_{\|}}\cdot {\boldsymbol x}_{_{\|}}(t')]}\,
         e^{-q_{_{\|}}\mid z'-z(t')\mid}}
         {{\left\vert\vphantom{\frac{1}{1}}\right.}_{{\boldsymbol r}={\boldsymbol r}(t)}}\, .
\end{eqnarray}
Let us digress a little bit and take a turn for a while. Equation (73), with the aid of Eq. (71), becomes
\begin{eqnarray}
{\boldsymbol F}=-\frac{e^2}{4\pi^2}\,&&\nabla \, \int d{\boldsymbol q}_{_{\|}}\, \int d\omega\,
e^{i({\boldsymbol q}_{_{\|}}\cdot {\boldsymbol x}_{_{\|}}-\omega t)}\,\frac{1}{q_{_{\|}}}\,\nonumber\\
&&\times \, \int dz'\,\left [\delta(z-z') + \smallint dz''\, V_{ee}(q_{_{\|}},z-z'')\,
               \chi({\boldsymbol q}_{_{\|}},\omega; z'', z')\right ] \nonumber\\
&&\times \,\int dt'\,{\large e^{i[\omega t'-{\boldsymbol q}_{_{\|}}\cdot {\boldsymbol x}_{_{\|}}(t')]}\,
         e^{-q_{_{\|}}\mid z'-z(t')\mid}}
         {{\left\vert\vphantom{\frac{1}{1}}\right.}_{{\boldsymbol r}={\boldsymbol r}(t)}}\, .
\end{eqnarray}
Another convenient way of writing Eq. (74), in its compact form, is as follows:
\begin{eqnarray}
{\boldsymbol F}=-\frac{e^2}{4\pi^2}\,&&\nabla \, \int d{\boldsymbol q}_{_{\|}}\, \int d\omega\,
e^{i({\boldsymbol q}_{_{\|}}\cdot {\boldsymbol x}_{_{\|}}-\omega t)}\,\frac{1}{q_{_{\|}}}\,\nonumber\\
&&\times \, \left [e^{-q_{_{\|}}(z-z_0)} - e^{q_{_{\|}}(z+z_0)}\,g(q_{_{\|}},\omega)\right ] \nonumber\\
&&\times \,\int dt'\,{\large e^{i[\omega t'-{\boldsymbol q}_{_{\|}}\cdot {\boldsymbol x}_{_{\|}}(t')]}\,
         e^{q_{_{\|}} v_z t'}}
         {{\left\vert\vphantom{\frac{1}{1}}\right.}_{{\boldsymbol r}={\boldsymbol r}(t)}}\, ,
\end{eqnarray}
where the symbol $g(...)$ related to the total DDCF $\chi(...)$ is a dimensionless quantity defined by
\begin{equation}
g(q_{_{\|}},\omega)=\frac{2\pi\,e^2}{\epsilon_b\,q_{_{\|}}}\int dz\int dz'\,
 e^{-q_{_{\|}}(z+z')}\, \chi({\boldsymbol q}_{_{\|}},\omega; z, z')\, .
\end{equation}
Some authors have termed $g(q_{_{\|}},\omega)$ a {\em surface response function} and have chosen to work with
Eq. (75) instead of Eq. (73) [38, 42].
We prefer to work with the induced force defined in terms of the inverse dielectric function, Eq. (73).  Let us
rewrite Eq. (73) in the form
\begin{eqnarray}
{\boldsymbol F}=-\frac{e^2}{4\pi^2}\,\nabla\, \int d{\boldsymbol q}_{_{\|}}\, \int d\omega\,
            e^{-i(\omega t- {\boldsymbol q}_{_{\|}}\cdot {\boldsymbol x}_{_{\|}})}\,\frac{1}{q_{_{\|}}}\,
            \int dz'\, e^{-q_{_{\|}}\mid z'-z_0 \mid}\,\epsilon^{-1}({\boldsymbol q}_{_{\|}}, \omega; z, z') \nonumber\\
\hspace{4.0cm}\times \,\int dt'\,e^{q_{_{\|}}z(t')}\,
             {\large e^{i[\omega t'-{\boldsymbol q}_{_{\|}}\cdot {\boldsymbol x}_{_{\|}}(t')]}}\,
         {{\left\vert\vphantom{\frac{1}{1}}\right.}_{{\boldsymbol r}={\boldsymbol r}(t)}}\, .
\end{eqnarray}
Decomposing the total force into parallel and perpendicular components yields: ${\boldsymbol F}={\boldsymbol F}_{\|}+\hat{z}\,F_z$.
We obtain
\begin{eqnarray}
{\boldsymbol F}_{\|}=-\frac{i\, e^2}{4\pi^2}\,\int d{\boldsymbol q}_{_{\|}}\, \int d\omega\,
             e^{-i(\omega - {\boldsymbol q}_{_{\|}}\cdot {\boldsymbol v}_{_{\|}})t}\,\frac{{\boldsymbol q}_{_{\|}}}{q_{_{\|}}}\,
          \int dz'\, e^{-q_{_{\|}}\mid z'-z_0 \mid}\,\epsilon^{-1}({\boldsymbol q}_{_{\|}}, \omega; z, z') \nonumber\\
\hspace{3.75cm}\times \,\int dt'\,e^{q_{_{\|}} v_z t'}\,
             {\large e^{i(\omega - {\boldsymbol q}_{_{\|}}\cdot {\boldsymbol v}_{_{\|}}) t'}}\,
         {{\left\vert\vphantom{\frac{1}{1}}\right.}_{{\boldsymbol z}=v_z t +z_0}}\, .
\end{eqnarray}
and
\begin{eqnarray}
{F}_{z}=-\frac{e^2}{4\pi^2}\,\int d{\boldsymbol q}_{_{\|}}\, \int d\omega\,
             e^{-i(\omega - {\boldsymbol q}_{_{\|}}\cdot {\boldsymbol v}_{_{\|}})t}\,\frac{1}{q_{_{\|}}}\,
\int dz'\, e^{-q_{_{\|}}\mid z'-z_0 \mid}\,{\partial_z}\epsilon^{-1}({\boldsymbol q}_{_{\|}}, \omega; z, z') \nonumber\\
\hspace{3.75cm}\times \,\int dt'\,e^{q_{_{\|}} v_z t'}\,
             {\large e^{i(\omega - {\boldsymbol q}_{_{\|}}\cdot {\boldsymbol v}_{_{\|}}) t'}}\,
         {{\left\vert\vphantom{\frac{1}{1}}\right.}_{{\boldsymbol z}=v_z t +z_0}}\, .
\end{eqnarray}
These are the exact results to be exploited for the purpose of calculating, e.g., the rate of energy loss, total
energy loss, stopping power, and/or the probability (or loss) function $P(q_{_{\|}}, \omega)$. Next we specify
the different configurations depending upon the orientation of the incoming electron with respect to the Q-2DEG
system [see Fig. 3].

\begin{figure}[htbp]
\includegraphics*[width=8cm,height=9cm]{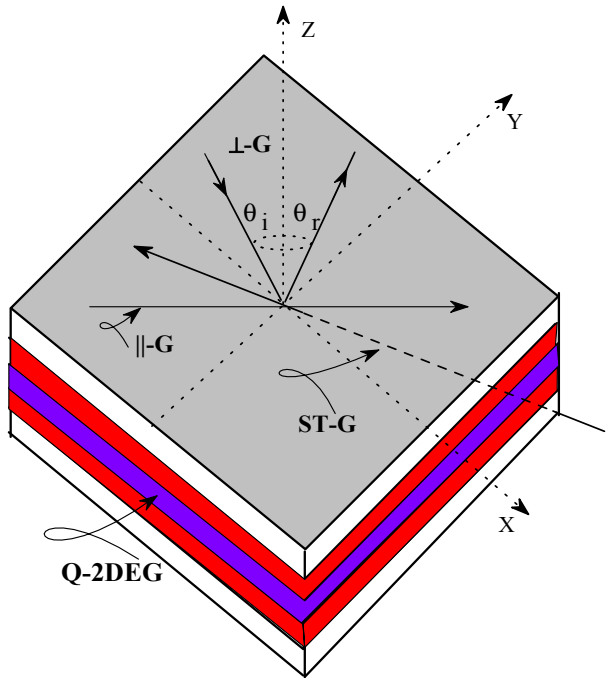}
\caption{(Color online) A schematic illustration of three principal configurations defining the geometry of the
electron beam with respect to the quasi-2DEG held in the structure. The symbolic terms $\|$-G, $\perp$-G, and
ST-G refer, respectively, to the parallel, perpendicular, and shooting-through configurations. The $\perp$-G,
strictly speaking, refers to the special case [$\theta_i=0=\theta_r$] of the specular reflection
[$\theta_i=\theta_r$].}
\label{fig3}
\end{figure}

\subsubsection{Parallel configuration}

By parallel configuration we mean the geometry where the incoming electron passes parallel to the x-y plane at a
distance, say, $z=z_0$ from the center of the Q-2DEG, which we presume to be the zero of the Cartesian coordinate
system. In this configuration, we would like to exploit Eq. (78) [and later Eq. (75)] in their compact form to
derive the rate of energy loss ($W'=dW/dt$), the total energy loss ($W$), and the loss function
[$P(q_{_{\|}}, \omega)$] for the system. Equation (60) together with Eq. (78) -- with proper limits $z(t)=z_0$,
and $v_z=0$ -- leaves us with
\begin{eqnarray}
\frac{dW}{dt}
&=&\frac{i\, e^2}{4\pi^2}\,\int d{\boldsymbol q}_{_{\|}}\, \int d\omega\,
             e^{-i(\omega - {\boldsymbol q}_{_{\|}}\cdot {\boldsymbol v}_{_{\|}})t}\,
            \frac{({\boldsymbol q}_{_{\|}}\cdot {\boldsymbol v}_{_{\|}})}{q_{_{\|}}}\nonumber\\
    && \times\,\int dz'\, e^{-q_{_{\|}}\mid z'-z \mid}\,\epsilon^{-1}({\boldsymbol q}_{_{\|}}, \omega; z, z')\,
       2\pi\,\delta (\omega - {\boldsymbol q}_{_{\|}}\cdot {\boldsymbol v}_{_{\|}})\,
         {{\left\vert\vphantom{\frac{1}{1}}\right.}_{\boldsymbol z=z_0}}\,\nonumber\\
&=&\frac{i\, e^2}{2\pi}\,\int d{\boldsymbol q}_{_{\|}} \int dz'\,
          \frac{({\boldsymbol q}_{_{\|}}\cdot {\boldsymbol v}_{_{\|}})}{q_{_{\|}}}\,
       e^{-q_{_{\|}}\mid z'-z_0 \mid}\,
  \epsilon^{-1}({\boldsymbol q}_{_{\|}}, \omega={\boldsymbol q}_{_{\|}}\cdot {\boldsymbol v}_{_{\|}}; z_0, z')\, .
\end{eqnarray}
This equation can also be cast in a different form to write
\begin{eqnarray}
\frac{dW}{dt}=\frac{i\, \epsilon_b}{4\pi^2}\,\int d{\boldsymbol q}_{_{\|}}\,
          ({\boldsymbol q}_{_{\|}} \cdot {\boldsymbol v}_{_{\|}}) \,
        V_s({\boldsymbol q}_{_{\|}}, \omega={\boldsymbol q}_{_{\|}}\cdot {\boldsymbol v}_{_{\|}}; z_0, z_0)\, .
\end{eqnarray}
where $V_s$ is the screened interaction potential energy of two test electrons occupying spatial sites $z=z_0$
and $z'=z_0$ [cf. Eq. (48)]. Since we are interested to compute the loss function $P(q_{_{\|}}, \omega)$, we
need to first determine the total energy loss. To this end, we integrate both sides of the first equality of
Eq. (80) with respect to the time. The result is
\begin{eqnarray}
W&=& e^2\, \int d{\boldsymbol q}_{_{\|}}\, \int d\omega\,
        \frac{({\boldsymbol q}_{_{\|}} \cdot {\boldsymbol v}_{_{\|}})}{q_{_{\|}}}\,
        [\delta (\omega - {\boldsymbol q}_{_{\|}}\cdot {\boldsymbol v}_{_{\|}})]^2\nonumber\\
&&\times\,\int dz'\, e^{-q_{_{\|}}\mid z'-z_0 \mid}\,
               {\rm Im}[\epsilon^{-1}({\boldsymbol q}_{_{\|}}, \omega; z_0, z')]\nonumber\\
&=& \int d{\boldsymbol q}_{_{\|}}\, \int d\omega\, \hbar\omega\, P(q_{_{\|}}, \omega)\, ,
\end{eqnarray}
where the loss function $P(q_{_{\|}}, \omega)$ is defined by
\begin{equation}
P(q_{_{\|}}, \omega)=\frac{e^2}{\hbar \omega}\,\frac{({\boldsymbol q}_{_{\|}} \cdot {\boldsymbol v}_{_{\|}})}{q_{_{\|}}}\,
   [\delta (\omega - {\boldsymbol q}_{_{\|}}\cdot {\boldsymbol v}_{_{\|}})]^2\,
   \int dz'\, e^{-q_{_{\|}}\mid z'-z_0 \mid}\,{\rm Im}[\epsilon^{-1}({\boldsymbol q}_{_{\|}}, \omega; z_0, z')]\, .
\end{equation}
This is the final expression for the loss function to be dealt with at the computational level. Next, if we use
Eq. (75) in conjunction with Eq. (60) -- with ${\boldsymbol x}_{_{\|}}(t)={\boldsymbol v}_{_{\|}}t$, $v_z=0$, and
$z=z_0 <0$ -- we obtain
\begin{eqnarray}
\frac{dW}{dt}&=&\frac{i\, e^2}{4\pi^2}\,\int d{\boldsymbol q}_{_{\|}}\, \int d\omega\,
            e^{-(\omega - {\boldsymbol q}_{_{\|}}\cdot {\boldsymbol v}_{_{\|}})t}\,
            \frac{({\boldsymbol q}_{_{\|}} \cdot {\boldsymbol v}_{_{\|}})}{q_{_{\|}}}\,\nonumber\\
&&\times\, \left [1 - e^{-2 q_{_{\|}}\mid z_0\mid}\,g(q_{_{\|}},\omega)\right ]\,
          2\pi\,\delta (\omega - {\boldsymbol q}_{_{\|}}\cdot {\boldsymbol v}_{_{\|}})\nonumber\\
&=&\frac{e^2}{2\pi}\,\int d{\boldsymbol q}_{_{\|}}\,
          \frac{({\boldsymbol q}_{_{\|}} \cdot {\boldsymbol v}_{_{\|}})}{q_{_{\|}}}\,
  e^{-2 q_{_{\|}}\mid z_0\mid}\,{\rm Im} [g(q_{_{\|}},\omega={\boldsymbol q}_{_{\|}}\cdot {\boldsymbol v}_{_{\|}})]\, .
\end{eqnarray}
This equation bears a formal resemblance with an equivalent expression in a layered system [42]. Making use of
the first equality of Eq. (84) and integrating it with respect to time yields
\begin{eqnarray}
W&=& e^2\, \int d{\boldsymbol q}_{_{\|}}\, \int d\omega\,
        \frac{({\boldsymbol q}_{_{\|}} \cdot {\boldsymbol v}_{_{\|}})}{q_{_{\|}}}\,
       [\delta (\omega - {\boldsymbol q}_{_{\|}}\cdot {\boldsymbol v}_{_{\|}})]^2\,
     e^{-2 q_{_{\|}}\mid z_0\mid}\,{\rm Im}[g(q_{_{\|}},\omega)]\nonumber\\
&=& \int d{\boldsymbol q}_{_{\|}}\, \int d\omega\, \hbar\omega\, P(q_{_{\|}}, \omega)\, ,
\end{eqnarray}
where the loss function $P(q_{_{\|}}, \omega)$ is defined by
\begin{equation}
P(q_{_{\|}}, \omega)=\frac{e^2}{\hbar \omega}\,\frac{({\boldsymbol q}_{_{\|}} \cdot {\boldsymbol v}_{_{\|}})}{q_{_{\|}}}\,
       [\delta (\omega - {\boldsymbol q}_{_{\|}}\cdot {\boldsymbol v}_{_{\|}})]^2\,
           e^{-2 q_{_{\|}}\mid z_0\mid}\,{\rm Im}[g(q_{_{\|}},\omega)]\, .
\end{equation}
One can thus see that there can be different ways to compute the rate of energy loss, the total energy loss, and
the loss function in these geometries. It is not difficult to prove that equating the right-hand sides (rhs) of
Eqs. (83) and (86) retrieves Eq. (76), just as expected.

\subsubsection{Perpendicular configuration}

As the name suggests, the perpendicular configuration refers to the geometry where the electron travels parallel
to the z direction and hits the Q-2DEG perpendicularly. This is the favorable geometry and is quite exploited in
the simpler surafce/interface and layered structures. It is very important to notice that this geometry requires
special attention with respect to the sign of $v_z$: we define the proper limits such that
\begin{eqnarray*}
z(t)=\left \{
\begin{array}{c}
+v_z t \,\,\,\,\, {\rm if \,\,\,\,\, t<0}\\
-v_z t \,\,\,\,\, {\rm if \,\,\,\,\, t>0}
\end{array}
\right . ,
\end{eqnarray*}
with $z_0=0$. In order to derive $W'$, $W$, and/or $P(...)$, we start with Eq. (79), which when substituted in
Eq. (60) yields
\begin{eqnarray}
\frac{dW}{dt}
&=&\frac{e^2\,v_z}{4\pi^2}\,\int d{\boldsymbol q}_{_{\|}}\, \int d\omega\,
             e^{-i(\omega - {\boldsymbol q}_{_{\|}}\cdot {\boldsymbol v}_{_{\|}})t}\,\frac{1}{q_{_{\|}}}\nonumber\\
&&\times\, \int dz'\, e^{-q_{_{\|}}z'}\,{\partial_z}\epsilon^{-1}({\boldsymbol q}_{_{\|}}, \omega; z, z') \nonumber\\
&&\hspace{0.0cm}\times \,\int dt'\,e^{q_{_{\|}} v_z t'}\,
             {\large e^{i(\omega - {\boldsymbol q}_{_{\|}}\cdot {\boldsymbol v}_{_{\|}}) t'}}\,
         {{\left\vert\vphantom{\frac{1}{1}}\right.}_{{\boldsymbol z}=v_z t}}\, .
\end{eqnarray}
Integrating both sides with respect to time gives the total energy loss defined by
\begin{eqnarray}
W&=&\frac{e^2\,v_z}{4\pi^2}\,\int d{\boldsymbol q}_{_{\|}}\, \int d\omega\,\frac{1}{q_{_{\|}}}\,
             \int dz'\, e^{-q_{_{\|}}z'}\nonumber\\
&& \times\, \int dt\, e^{-i(\omega - {\boldsymbol q}_{_{\|}}\cdot {\boldsymbol v}_{_{\|}})t}\,
          {\partial_z}\epsilon^{-1}({\boldsymbol q}_{_{\|}}, \omega; z, z') \nonumber\\
&&\hspace{0.0cm}\times \,\int dt'\,e^{q_{_{\|}} v_z t'}\,
             {\large e^{i(\omega - {\boldsymbol q}_{_{\|}}\cdot {\boldsymbol v}_{_{\|}}) t'}}\,
         {{\left\vert\vphantom{\frac{1}{1}}\right.}_{{\boldsymbol z}=v_z t}}\, .
\end{eqnarray}
It is not difficult to simplify the integrals in the second and third lines of this equation, provided that we
take proper care of how $v_z$ changes the sign at $t=0$. We obtain
\begin{eqnarray}
W&=&\frac{i e^2}{2\pi^2}\,\frac{\alpha}{\beta}\,\int d{\boldsymbol q}_{_{\|}}\, \int d\omega\,
             \int dz\, e^{-q_{_{\|}}z}\nonumber\\
&& \times\, \int dz'\, e^{-i\alpha (z'/v_z)}\,\epsilon^{-1}({\boldsymbol q}_{_{\|}}, \omega; z', z)\nonumber\\
&=& \int d{\boldsymbol q}_{_{\|}}\, \int d\omega\, \hbar\omega\, P(q_{_{\|}}, \omega)\, ,
\end{eqnarray}
where the loss function $P(q_{_{\|}}, \omega)$ is now defined by
\begin{eqnarray}
P(q_{_{\|}}, \omega)
      &=&\frac{e^2}{2\pi^2\hbar}\,\frac{\alpha}{\omega}\,\frac{1}{\beta}\,
        \int dz\, e^{-q_{_{\|}}z}\,\int dz'\, e^{-i\alpha (z'/v_z)}\,
            {\rm Im}[\epsilon^{-1}({\boldsymbol q}_{_{\|}}, \omega; z', z)]\, .
\end{eqnarray}
and the substitution $\alpha=(\omega-{\boldsymbol q}_{_{\|}} \cdot {\boldsymbol v}_{_{\|}})$ and $\beta=[\alpha^2+(q_{_{\|}}v_z)^2]$.
Next, if we employ Eq. (75) [with $\nabla_z=\frac{1}{v_z}\frac{\partial}{\partial t}$] along with Eq. (60) -- with
${\boldsymbol x}_{_{\|}}(t)={\boldsymbol v}_{_{\|}}t$, $z(t)=v_z t$, and $z_0=0$ -- we obtain the total energy loss
given by
\begin{eqnarray}
W&=&\frac{i e^2}{4\pi^2}\,
           \int d{\boldsymbol q}_{_{\|}}\, \int d\omega\,\frac{\alpha}{q_{_{\|}}}\,g(q_{_{\|}},\omega)\,\nonumber\\
&& \times\, \int dt\,
           {\large e^{-i(\omega - {\boldsymbol q}_{_{\|}}\cdot {\boldsymbol v}_{_{\|}}) t}}\,e^{q_{_{\|}} v_z t}\,
\int dt'\,
{\large e^{i(\omega - {\boldsymbol q}_{_{\|}}\cdot {\boldsymbol v}_{_{\|}}) t'}}\,e^{q_{_{\|}} v_z t'}\nonumber\\
&=&\frac{e^2\,v^2_z}{\pi^2}\,\int d{\boldsymbol q}_{_{\|}}\, \int d\omega\,\,
                \frac{q_{_{\|}}\alpha}{\beta^2}\,\, {\rm Im}[g(q_{_{\|}},\omega)]\nonumber\\
&=& \int d{\boldsymbol q}_{_{\|}}\, \int d\omega\,\hbar\omega\, P(q_{_{\|}}, \omega)\, .
\end{eqnarray}
An equivalent expression was derived within the classical approach for the dielectric superlattices [39]. The loss
function $P(q_{_{\|}}, \omega)$ in Eq. (91) is given by
\begin{equation}
P(q_{_{\|}}, \omega)=\frac{e^2}{\pi^2\,\hbar}\,\frac{\alpha}{\omega}\,\frac{q_{_{\|}}\,v^2_z}{\beta^2}\,\,
               {\rm Im}[g(q_{_{\|}},\omega)]
\end{equation}
Equating the rhs of Eqs. (90) and (92)
leads us to recover Eq. (76). There is a great advantage in deriving and computing the loss function
$P(q_{_{\|}}, \omega)$ while studying the EELS. Not only do we retain the whole system response intact in it, a very substantial saving in computational time is achieved by getting rid of the two integrals over ${\boldsymbol q}_{_{\|}}$
and over $\omega$.

\subsubsection{Shooting-through configuration}

In order to describe this geometry, the initial step is to calculate the total work done by the induced force --
comprised of the parallel plus the perpendicular components -- to obtain the total energy loss the incoming
electron suffers from. Just as before, we start with Eq. (60) together with Eqs. (78) and (79) to write
\begin{eqnarray}
W&=&\frac{e^2}{4\pi^2}\,\int d{\boldsymbol q}_{_{\|}}\, \int d\omega\nonumber\\
&&\times\,\left [\frac{i{\boldsymbol q}_{_{\|}}\cdot {\boldsymbol v}_{_{\|}}}{q_{_{\|}}}\,
          \int dt\, e^{-i(\omega - {\boldsymbol q}_{_{\|}}\cdot {\boldsymbol v}_{_{\|}})t}\,
         \int dz'\, e^{-q_{_{\|}}z'}\,\epsilon^{-1}({\boldsymbol q}_{_{\|}}, \omega; z, z')\right. \nonumber\\
&&\hspace{3.75cm}\times\,\int dt'\,e^{q_{_{\|}} v_z t'}\,
             {\large e^{i(\omega - {\boldsymbol q}_{_{\|}}\cdot {\boldsymbol v}_{_{\|}}) t'}}\nonumber\\
&& \,\,\,\,\,\,\,\,\,\,\,+ \frac{v_z}{q_{_{\|}}}\,
          \int dt\, e^{-i(\omega - {\boldsymbol q}_{_{\|}}\cdot {\boldsymbol v}_{_{\|}})t}\,
         \int dz'\, e^{-q_{_{\|}}z'}\,{\partial_z}\epsilon^{-1}({\boldsymbol q}_{_{\|}}, \omega; z, z') \nonumber\\
&&\hspace{3.75cm}\left.\times\,\int dt'\,e^{q_{_{\|}} v_z t'}\,
             {\large e^{i(\omega - {\boldsymbol q}_{_{\|}}\cdot {\boldsymbol v}_{_{\|}}) t'}}\right ]\,
         {{\left\vert\vphantom{\frac{1}{1}}\right.}_{z=v_z t}}\, .
\end{eqnarray}
Simplifying $[{\partial_z}\epsilon^{-1}({\boldsymbol q}_{_{\|}}, \omega; z, z')]$ and using $z=v_z t$ leaves us with
\begin{eqnarray}
W&=&\frac{e^2}{4\pi^2}\,\int d{\boldsymbol q}_{_{\|}}\, \int d\omega\nonumber\\
&&\times\,\left [\frac{i{\boldsymbol q}_{_{\|}}\cdot {\boldsymbol v}_{_{\|}}}{q_{_{\|}}v_z}\,
          \int dz'\, e^{-q_{_{\|}}z'}\,
          \int dz''\, e^{-i(\omega - {\boldsymbol q}_{_{\|}}\cdot {\boldsymbol v}_{_{\|}})z''/v_z}\,
                   \epsilon^{-1}({\boldsymbol q}_{_{\|}}, \omega; z'', z')\right. \nonumber\\
&&\hspace{3.75cm}\times\,\int dt'\,e^{q_{_{\|}} v_z t'}\,
             {\large e^{i(\omega - {\boldsymbol q}_{_{\|}}\cdot {\boldsymbol v}_{_{\|}}) t'}}\nonumber\\
&& \,\,\,\,\,\,\,\,\,\,\,\,+ \frac{i\alpha}{q_{_{\|}}v_z}\,
          \int dz'\, e^{-q_{_{\|}}z'}\,
          \int dz''\, e^{-i(\omega - {\boldsymbol q}_{_{\|}}\cdot {\boldsymbol v}_{_{\|}})z''/v_z}\,
                   \epsilon^{-1}({\boldsymbol q}_{_{\|}}, \omega; z'', z') \nonumber\\
&&\hspace{3.75cm}\left.\times\,\int dt'\,e^{q_{_{\|}} v_z t'}\,
             {\large e^{i(\omega - {\boldsymbol q}_{_{\|}}\cdot {\boldsymbol v}_{_{\|}}) t'}}\right ]\,
\end{eqnarray}
Summing carefully the two terms inside the square brackets piecewise yields
\begin{eqnarray}
W&=&\frac{i\,e^2}{4\pi^2\,v_z}\,\int d{\boldsymbol q}_{_{\|}}\, \int d\omega\,\, \frac{\omega}{q_{_{\|}}}\,
          \int dz'\, e^{-q_{_{\|}}z'}\,
\int dt'\,e^{q_{_{\|}} v_z t'}\,
             {\large e^{i(\omega - {\boldsymbol q}_{_{\|}}\cdot {\boldsymbol v}_{_{\|}}) t'}}\nonumber\\
&&\hspace{3.25cm}\times\,\int dz''\, e^{-i(\omega - {\boldsymbol q}_{_{\|}}\cdot {\boldsymbol v}_{_{\|}})z''/v_z}\,
                   \epsilon^{-1}({\boldsymbol q}_{_{\|}}, \omega; z'', z')\nonumber\\
&=&\frac{i\,e^2}{2\pi^2\,\beta}\,\int d{\boldsymbol q}_{_{\|}}\, \int d\omega\,\, \omega \,
          \int dz'\, e^{-q_{_{\|}}z'}\nonumber\\
&&\hspace{3.25cm}\times\,\int dz''\, e^{-i(\omega - {\boldsymbol q}_{_{\|}}\cdot {\boldsymbol v}_{_{\|}})z''/v_z}\,
                   \epsilon^{-1}({\boldsymbol q}_{_{\|}}, \omega; z'', z')\nonumber\\
&=& \int d{\boldsymbol q}_{_{\|}}\, \int d\omega\, \hbar\omega\, P(q_{_{\|}}, \omega)\, ,
\end{eqnarray}
where the loss function $P(q_{_{\|}}, \omega)$ is now defined by
\begin{eqnarray}
P(q_{_{\|}}, \omega)
          =\frac{e^2}{2\pi^2\,\hbar}\,\frac{1}{\beta}\,
             \int dz\, e^{-q_{_{\|}}z}\,\int dz'\, e^{-i\alpha (z'/v_z)}\,
                 {\rm Im}[\epsilon^{-1}({\boldsymbol q}_{_{\|}}, \omega; z', z)]\, .
\end{eqnarray}
Next, if we employ Eq. (75) [with $\nabla_z=\frac{1}{v_z}\frac{\partial}{\partial t}$] along with Eq. (60) -- with
${\boldsymbol x}_{_{\|}}(t)={\boldsymbol v}_{_{\|}} t$ and $z(t)=v_z t$ -- we obtain the total energy loss given by
\begin{eqnarray}
W&=&\frac{e^2}{4\pi^2}\,\int d{\boldsymbol q}_{_{\|}}\, \int d\omega\,\, \frac{\omega}{q_{_{\|}}}\,
        {\rm Im}[g(q_{_{\|}},\omega)]\nonumber\\
&&\times\,\int dt\,e^{q_{_{\|}} v_z t}\,{\large e^{-i\alpha t}}\,
         \int dt'\,e^{q_{_{\|}} v_z t'}\,{\large e^{-i\alpha t'}}\nonumber\\
&=& \frac{e^2v^2_z}{\pi^2}\,\int d{\boldsymbol q}_{_{\|}}\, \int d\omega\,
        \frac{\omega q_{_{\|}}}{\beta^2}\, {\rm Im}[g(q_{_{\|}},\omega)]\nonumber\\
&=& \int d{\boldsymbol q}_{_{\|}}\, \int d\omega\, \hbar\omega\, P(q_{_{\|}}, \omega)\, ,
\end{eqnarray}
where the loss function $P(q_{_{\|}}, \omega)$ is defined as follows.
\begin{equation}
P(q_{_{\|}}, \omega)=\frac{e^2}{\pi^2\,\hbar}\,\frac{q_{_{\|}}\,v^2_z}{\beta^2}\,
                       \,{\rm Im}[g(q_{_{\|}},\omega)]
\end{equation}
Again, equating the rhs of Eqs. (96) and (98) regains the correlation in Eq. (76).
There are certain things we need to recognize regarding the computation of $W'$, $W$, or $P(...)$ irrespective
of the preference for the geometry. The prefactors do not matter much because they can at the most influence
the height, width, or (sometimes a little) shape of the loss peaks, but not the {\em position} in energy. What
matters most is the system response that comes from the integration of the factor that involves the inverse
dielectric function [or the surface response function, as the case may be]. Equations (83), (86), (90), (92),
(96), and (98) are the final results of this section. A few further analytical details needed for the purpose of
computation will be given and briefly discussed later in Sec. II.G.

\subsection{The inelastic light scattering}

When light is scattered from a medium, most of the photons are scattered elastically, i.e., where the scattered
photons have the same energy (and hence wavelength) as the incident photons. However, a small fraction of light
($\approx 1$ in $10^7$ photons) is scattered inelastically, i.e., where the scattered and incident photons differ
in energy (and hence in wavelength). The process leading to this inelastic scattering is known the Raman effect
or Raman scattering [after the discoverer Sir C. V. Raman]. The difference in energy between the incident and
scattered photons is equal to the energy of the (respective) excitations of the medium.

In solids, there can be various types of the elementary excitations such as phonons, magnons, plasmons, ...etc.
Historically, the scattering by optical (acoustic) phonons is called Raman (Brillouin) scattering. For the
electronic Raman scattering, often the term inelastic light scattering (ILS) is used. The creation (destruction)
of the excitations during the scattering process is called Stokes (anti-Stokes) process within the medium. Each
of the scattered photons in the Stokes (anti-Stokes) process is associated with a gain (loss) in energy: $\hbar\omega_s=\hbar\omega_i \mp \hbar\omega$, where the minus (plus) sign stands for the Stokes (anti-Stokes)
process; subscript i (s) refers to the incident (scattered) photons. The conservation of momentum then requires
that $k_s=k_i\mp q$. Here $\hbar\omega$ ($q$) is the energy (momentum) of the elementary excitation in the medium.

In semiconducting structures (as is the case here), the energies of the elementary excitations are, generally,
smaller than the energy of the incident laser light (i.e., $\hbar\omega < \hbar\omega_i$). A particular strength
of the ILS therefore is that elementary excitations with energies in the FIR spectral range can be measured in
the visible range, where powerful lasers and detectors are available.  A further strength of the ILS is the
possibility to transfer a finite quasi-momentum ($q$) to the excitation during the scattering process. The maximum
$q$ can be transferred, e.g., by employing the exact back-scattering geometry (BSG) [i.e., where the directions of
the incident and scattered light are antiparallel]. In the BSG, $q_{max}$ is twice the momentum of light (with the
assumption that $\lambda_i=\lambda_s$): $q_{max}=2 (2\pi/\lambda_i)$.

Since the Raman scattering is such a powerful optical spectroscopy to study the interacting electron systems in the
quasi-n dimensions (with $n=3$, 2, 1, 0) -- including, e.g., the quantum Hall systems -- we thought it worthwhile
to embark on a systematic and thorough formulation of the process of ILS. The coupling of the electromagnetic (EM)
radiation with the electron system is taken into account by replacing the momentum $\boldsymbol p$ of the electron
with ${\boldsymbol p}+(e/c){\boldsymbol A}$ [with the fundamental electronic charge defined as $-e$, with $e>0$] in
the Hamiltonian $H_0$ of the unperturbed system. Here ${\boldsymbol A}$ is the sum of the vector potentials of the
incident and scattered EM fields.  One of the virtues of our treatment is that we will deal with the problem without
making any specific approximation regarding the single-particle eigenstates $\mid \alpha>$:
$H_0\mid \alpha>=\epsilon_{\alpha}\mid \alpha>$. Here $\alpha\equiv (n, k, \sigma)$; $n$, $k$, and $\sigma$ being,
respectively, the band index, the wave vector, and the spin index. The term {\em band} refers to the (conventional)
{\em valence} and {\em conduction} bands and is not a substitution for the {\em subband}. It is also noteworthy
that we will proceed throughout as if we are dealing with a conventional (3D) system: the specific dimension of
the system will not be an issue until finally we calculate the correlation (or response) function of the system of
interest. The carrier density operator $n({\boldsymbol r})=\Psi^+({\boldsymbol r})\Psi({\boldsymbol r})$, where the
field operators, in the second quantization, can be written as
\begin{eqnarray}
\Psi({\boldsymbol r})&=&\sum_{n {\boldsymbol k} \sigma}\,c_{n {\boldsymbol k} \sigma}\,
           \mid n; {\boldsymbol k}, \sigma \left >\right. \nonumber\\
\Psi^+({\boldsymbol r})&=&\sum_{n {\boldsymbol k} \sigma}\,c^+_{n {\boldsymbol k} \sigma}\,
           \left < n; {\boldsymbol k}, \sigma \mid \right.\, ,
\end{eqnarray}
where $c^{+}_x$ ($c_x$) refers to the creation (destruction) operator for the conduction electrons and they satisfy
the anticommutation relations for fermions. The EM field is assumed to interact with the system as an ensemble of
harmonic oscillators and its vector potential ${\boldsymbol A}({\boldsymbol r}, t)$ can be expressed as [64]
\begin{equation}
{\boldsymbol A}({\boldsymbol r}, t)=\sum_{\mbox{\boldmath$\eta$} \lambda}\,
             \left (\frac{2\pi \hbar c^2}{V\omega(\mbox{\boldmath$\eta$})} \right )^{1/2}\,
 \left [a_{\mbox{\boldmath$\eta$} \lambda}\,e^{i(\mbox{\boldmath $\eta$}\cdot {\boldsymbol r}-\omega_i t)}
             + H.c.\right ]\,
                    \hat {e}(\mbox{\boldmath$\eta$} \lambda)\, ,
\end{equation}
where $a^+_{\mbox{\boldmath$\eta$} \lambda}$ ($a_{\mbox{\boldmath$\eta$} \lambda}$) is the creation (destruction)
operator for a photon of wave vector $\mbox{\boldmath$\eta$}$, polarization index $\lambda$, unit polarization vector
$\hat {e}(\mbox{\boldmath$\eta$} \lambda)$, and frequency $\omega (\mbox{\boldmath$\eta$})$. Here $V$ is the volume
of the system and $\omega_i$ is the frequency of the incident light. We shall assume that the EM field in the system
is purely transverse so that the Coulomb gauge reigns implying that $\nabla\cdot{\boldsymbol A}=0$ and
$\mbox{\boldmath$\eta$}\cdot \hat {e}(\mbox{\boldmath$\eta$} \lambda)=0$. The interaction Hamiltonian which describes
the interaction between the electrons and the EM radiation is given by
\begin{equation}
H_{int}=H_1 +H_2\, ,
\end{equation}
where
\begin{equation}
H_1 = \frac{e}{mc}\,\int d{\boldsymbol r}\, n({\boldsymbol r})\,{\boldsymbol p}\cdot{\boldsymbol A}
           ({\boldsymbol r}, t)
\end{equation}
\begin{equation}
H_2 = \frac{e^2}{2mc^2}\,\int d{\boldsymbol r}\, n({\boldsymbol r})\,{\boldsymbol A}
                 ({\boldsymbol r}, t)\cdot{\boldsymbol A}({\boldsymbol r}, t)
\end{equation}
and ${\boldsymbol p}$ is the momentum operator. The term $H_1$ ($H_2$) is clearly linear (quadratic) in the vector
potential ${\boldsymbol A}({\boldsymbol r}, t)$. It turns out that $H_1$ ($H_2$) gives the dominant contribution
to interband (intraband) transitions. A simple argument to reveal this owes to P.M. Platzman and is succinctly
discussed here. It is based on the fact that if the electron-radiation interaction is treated within the
perturbation theory, $H_1$ ($H_2$) needs to be carried out to second (first) order for the two-photon processes of
interest in the ILS experiments [see Fig. 4 for a brief account of the process]. The second-order matrix element of
$H_1$ has the schematic form
\begin{figure}[htbp]
\includegraphics*[width=8cm,height=6.5cm]{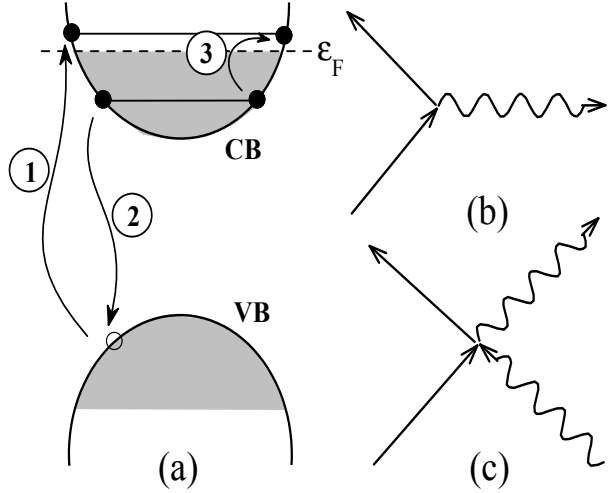}
\caption{(a) A schematic representation of the two-step process: in step 1, the incident photon excites an electron
in the valence band (VB) into an excited state (above the Fermi level) in the conduction band (CB) leaving a hole
in the VB; in step 2, an electron from the CB (below the Fermi level) recombines with the hole in the VB emitting
an outgoing photon shifted both in energy and momentum. As a net effect,  an elementary electronic excitation is
created inside the CB through intermediate VB (IVB) states. The nonresonant scheme neglects the IVB states and
approximates the RRS process to take place entirely within the CB as depicted by step 3 in the picture. In a
nut-shell, nonresonant process is not equivalent to the RRS process: the latter involving steps 1 and 2 depends
explicitly on the incident-photon energy, while the former referred to step 3 depends only on the energy difference
between the incident and scattered photons. This difference turns out to be crucial in the RRS theory, particularly
with respect to the extra SPE-like peaks observed in the experiment. In the right panel, (b) and (c) are the Feynman
diagrams for the light scattering via ${\boldsymbol p}\cdot{\boldsymbol A}$ and ${\boldsymbol A}\cdot{\boldsymbol A}$
terms, respectively. The solid (wiggly) line refers to the electron (photon) Green function.}
\label{fig4}
\end{figure}
\begin{equation}
P(H_1) \approx \left (\frac{e}{mc} \right )^2\,\frac{({\boldsymbol p}\cdot{\boldsymbol A})^2}{\Delta\epsilon}
\end{equation}
and the first-order matrix element of $H_2$ has the schematic form
\begin{equation}
P(H_2) \approx \left (\frac{e^2}{mc^2} \right )\,{\boldsymbol A}^2
\end{equation}
where $\Delta\epsilon$ is the sum or difference of an excitation energy of the many-electron energy system and a
typical photon energy. In the usual circumstances, the photon energy is much larger than the excitation energy, and
one can safely approximate $\Delta\epsilon$ by the photon energy $\hbar\omega$. Then the ratio $\mathbb{R}\equiv P(H_1)/P(H_2)$ is given by
\begin{equation}
\mathbb{R}\,\approx\, \frac{p^2}{m\,\Delta\epsilon}\,\approx\,\frac{p^2}{m\hbar\omega}
\end{equation}
Next, if $p\approx \hbar\omega/c$, then Eq. (106) assumes the form
\begin{equation}
\mathbb{R}\,\approx\, \frac{\hbar\omega}{mc^2}
\end{equation}
Since $\hbar\omega << mc^2$, the ratio $\mathbb{R}$ is completely negligible for the usual experiments, so we conclude
that the $H_2$ term remains dominant in the intraband transitions. For the interband case, the matrix elements of $H_2$ become negligibly small and in fact vanish as $\mbox{\boldmath$\eta$} \rightarrow 0$ due to the orthogonality of the electronic wave functions. The matrix elements of $H_1$ are, on the other hand, large for direct allowed transitions
between energy bands, because the matrix elements of ${\boldsymbol p}$ are large. Thus the $H_1$ term becomes dominant
in the interband case. An important question, however is: when, where, and why do the interband transitions really
matter (more than or equal to the intraband transitions)? We intend to address this question in the end of this section
in order not to digress from our principal theme. As such, we begin to focus on the theory of the non-resonant ILS experiments, i.e., neglecting the interband transitions. The process we are concerned with is the absorption of an
incident photon ($\omega_i, \mbox{\boldmath$\eta$}_i$) and the emission of a scattered photon ($\omega_s, \mbox{\boldmath$\eta$}_s$). The radiation state vectors are characterized by the occupation numbers of the incident and scattered photons, $n_i$ and $n_s$, respectively: $\mid R \left >\right.=\mid n_i, n_s \left >\right.$. The radiation
state vectors are: the initial state $\mid R i\left >\right.=\mid n_i, 0 \left >\right.$ and the final state
$\mid R f\left >\right.=\mid n_i-1, 1\left >\right.$. The evaluation of the matrix elements of $H_2$ between the states
$\mid R f\left >\right.$ and $\mid R i\left >\right.$ is carried out by making use of Eqs. (100) and (103) and the
properties of boson creation and destruction operators. Notice, the operators which are the products of two fermion
operators have the character of boson operators. The result, within the effective-mass approximation (EMA) [65], is
\begin{equation}
\left <\right. R f\mid H_2\mid R i \left >\right. =
   \frac{2\pi\hbar e^2}{m^*V}\,\Big (\frac{n_i}{\omega_i\omega_s} \Big )^{1/2}\,(\hat {e}_i \centerdot \hat{e}_s)\,
\int d{\boldsymbol r}\, n({\boldsymbol r})\,
                      e^{i(\mbox{\boldmath$\eta$}_i-\mbox{\boldmath$\eta$}_s){\mbox{$\centerdot$}} {\boldsymbol r}}\, ,
\end{equation}
where $m^*$ is the effective mass of the free carriers. Since we are dealing only with intraband transitions, the sums
over the band index $n$ can be suppressed in Eqs. (99) -- restricting them to only conduction (valence) band states
depending upon whether the semiconductor is n- (p-) type. Also, we can further simplify the notation by combining the
spin index with the wave vector, since the two always go together. If we substitute
${\boldsymbol q}=\mbox{\boldmath$\eta$}_i-\mbox{\boldmath$\eta$}_s$, then
\begin{eqnarray}
n_{\boldsymbol q}&=&\frac{1}{V}\,\int d{\boldsymbol r}\, n({\boldsymbol r})\,
             e^{i{\boldsymbol q}{\mbox{$\centerdot$}} {\boldsymbol r}}\nonumber\\
&=&\frac{1}{V}\,\sum_{\boldsymbol k}\,c^+_{\boldsymbol {k+q}}\,c_{\boldsymbol k}
\end{eqnarray}
can be interpreted as the Fourier transform of $n({\boldsymbol r})$ implying that the matrix element, Eq. (108), is
simply proportional to the $n_{\boldsymbol q}$. The second equality is obtained by using Eqs. (99) in the definition
of $n(\boldsymbol r)$ within the EMA. Substituting Eq. (109) into Eq. (108) yields
\begin{equation}
{M_2}\equiv \left < R f\mid H_2\mid R i \right > =
   \frac{2\pi\hbar e^2}{m^*V}\,\Big (\frac{n_i}{\omega_i\omega_s} \Big )^{1/2}\,(\hat {e}_i \centerdot \hat{e}_s)\,
                 \sum_{\boldsymbol k}\,c^+_{\boldsymbol{k+q}}\,c_{\boldsymbol k} \, .
\end{equation}
Next, we will calculate the transition probability $W$ using the Fermi Golden rule. For this, we first need the matrix elements of $M_2$ with respect to the free carrier states: $\left < Cf \mid M_2 \mid Ci\right >=
\left < Cf \mid \left < R f\mid H_2\mid R i \right > \mid Ci\right >$. It thus becomes clear that here $R$ ($C$) refers
to the radiation (carrier) states. The Golden rule then allows us to write
\begin{equation}
W=\frac{2\pi}{\hbar}\,\sum_{f}\,{\big |} \big < Cf {|} M_2 {|} Ci\big > {\big |}^2\, \delta(\epsilon_f-\epsilon_i)\, ,
\end{equation}
where $\epsilon_i$ ($\epsilon_f$) is the energy of the initial (final) state of the whole system (i.e., electrons plus photons). If the eigenenergies of the free carrier Hamiltonian $H_0$ are designated by $\hbar \omega_k$, then
\begin{eqnarray}
\epsilon_f -\epsilon_i
&=& \hbar\omega_{k_{f}} + \hbar \omega_{s} - \hbar\omega_{k_{i}} -\hbar\omega_{i}\nonumber\\
&=& \hbar\omega_{k_{f}} - \hbar\omega_{k_{i}} - \hbar\omega \, ,
\end{eqnarray}
where $\omega=\omega_i - \omega_s$. Since we have neglected the spin-orbit coupling the present system is free from any
cause and effect of the radiation-induced transition on the spin state. Next, we make use of the integral representation
of the Dirac delta function to write
\begin{eqnarray}
\delta(\epsilon_f - \epsilon_i)
&=& \frac{1}{2\pi\hbar}\, \int dt\, e^{i(\omega +\omega_{k_{i}} - \omega_{k_{f}})t}\, .
\end{eqnarray}
Substituting Eq. (113) into Eq. (111) redefines the transition probability as 
\begin{equation}
W=\frac{1}{\hbar^2}\, \sum_{f}\,\int dt \, e^{i\omega t}\,
          \big < Ci \mid M^+_2(t) \mid Cf\big > \big < Cf \mid M_2(0) \mid Ci\big >\, ,
\end{equation}
where $\left [{\rm using} H_{0}{\big |} k\big >=\epsilon_{k}{\big |} k\big >, {\rm with} \mid Ci\big >=\mid k_i\big >
{\rm and} \mid Cf\big >=\mid k_f\big >\right ]$
\begin{equation}
M^{+}_{2}(t)=e^{iH_{0} t/\hbar}M^{+}_{2}(0)e^{-iH_{0} t/\hbar}\, .
\end{equation}
Making use of the closure relation satisfied by the eigenfunctions, one can carry out the sum over $f$ and then take a
thermal average over grand canonical ensemble. Thus the average transition probability is defined as
\begin{equation}
\big < W \big >=\frac{1}{\hbar^2}\, \int dt \, e^{i\omega t}\,\big < M^+_2(t) M_2(0) \big >\, ,
\end{equation}
where the angular brackets now indicate a thermal average. The number of scattered photon states (of given polarization)
in a solid angle $d\Omega$ and the frequency interval $d\omega$ is given by
\begin{eqnarray}
N_p
=\frac{V}{(2\pi)^2}\,\eta^2_s\,\frac{d\eta_s}{d\omega_s}\,d\omega\,d\Omega
=\frac{V}{(2\pi)^2}\,\frac{\omega^2_s}{c^3}\,d\omega\,d\Omega\, ,
\end{eqnarray}
where the relation $\eta_s=\omega_s/c$ leads to the second equality; $c$ is the speed of light in vacuum. The number
of transitions per unit time, per unit solid angle, per unit frequency interval is given by
\begin{eqnarray}
\frac{d^2 \big < W \big >}{d\omega d\Omega}
=\frac{\big < W \big > N_{p}}{d\omega d\Omega}
=\frac{V \omega^2_s}{(2\pi)^3\hbar^2 c^3}\, \int dt \, e^{i\omega t}\,\big < M^+_2(t) M_2(0) \big >\, .
\end{eqnarray}
Dividing this by the area $A$ of the sample illuminated by the incident beam and by the flux of the incident photons
$n_i c/V$ gives the light scattering efficiency defined by
\begin{eqnarray}
\frac{d^2 S}{d\omega d\Omega}
=\frac{V^2 \omega^2_s}{(2\pi)^3\hbar^2 c^4 n_i A}\, \int dt \, e^{i\omega t}\,\big < M^+_2(t) M_2(0) \big >\, .
\end{eqnarray}
Making use of the expression for $M_2$ in terms of $n_{\boldsymbol q}$ [see Eqs. (108) - (110)] leads us to obtain
\begin{eqnarray}
\frac{d^2 S}{d\omega d\Omega}
=\frac{V^2}{2\pi A}\, r^2_0\, \Big (\frac{\omega_s}{\omega_i}\Big )\,(\hat {e}_i \centerdot \hat{e}_s)^2\,
          \int dt \, e^{i\omega t}\,\big < n^+_{\boldsymbol q}(t) n_{\boldsymbol q}(0) \big >\, ,
\end{eqnarray}
where $r_0=e^2/m^* c^2$ is the classical electron radius, but with an effective mass $m^*$. Thus the scattering
efficiency is proportional to the Fourier transform of the DDCF at wave vector $\boldsymbol q$. Eq. (120) is, in
a sense, the analogue of the fluctuation-dissipation theorem for the ILS. Let us make use of the definition of $n_{\boldsymbol q}$ [see Eq. (109)] to write Eq. (120) in the form
\begin{eqnarray}
\frac{d^2 S}{d\omega d\Omega}
=\frac{1}{2\pi A}\, r^2_0\, \Big (\frac{\omega_s}{\omega_i}\Big )\,(\hat {e}_i \centerdot \hat{e}_s)^2\,
         \sum_{{\boldsymbol k}{\boldsymbol k}'}\, \int dt \, e^{i\omega t}\,
         \big < c^+_{\boldsymbol {k{\mbox{-}}q}}(t)\,c_{\boldsymbol {k}}(t)\,
         c^+_{\boldsymbol {k'+q}}(0)\,c_{\boldsymbol {k'}}(0) \big >\, ,
\end{eqnarray}
where we have used $n^+_{\boldsymbol q}=n_{-{\boldsymbol q}}$ which follows from the fact that $n(\boldsymbol r)$ is
real. If we exploit the periodicity of the energy bands in the $\boldsymbol k$-space, we can replace
${\boldsymbol k}'$ by $\boldsymbol {k'-q}$ in Eq. (121). In addition, multiplying Eq. (121) by the area $A$ of the
illuminated sample yields the differential scattering cross-section defined by
\begin{eqnarray}
\frac{d^2 \sigma}{d\omega d\Omega}
=\frac{1}{2\pi}\, r^2_0\, \Big (\frac{\omega_s}{\omega_i}\Big )\,(\hat {e}_i \centerdot \hat{e}_s)^2\,
         \sum_{{\boldsymbol k}{\boldsymbol k}'}\, \int dt \, e^{i\omega t}\,
         \big < c^+_{\boldsymbol {k{\mbox{-}}q}}(t)\,c_{\boldsymbol {k}}(t)\,
         c^+_{\boldsymbol {k'}}(0)\,c_{\boldsymbol {k'{\mbox{-}}q}}(0) \big >\, ,
\end{eqnarray}
Some variants of this result can be found in the classic works in the late sixties [see, e.g., Refs. 46-51]. Next,
we evaluate the Fourier-transformed correlation function (inside the integrand) by making use of the double-time
retarded Green functions. To do this, we define the free-carrier Hamiltonian in the conduction band in the second
quantization given by
\begin{equation}
H=\sum_{k}\,\epsilon_k\, c^+_k c_k + \frac{1}{2}\,\sum_{kk'}\,\sum_q\,V_{q}\,
c^+_{\boldsymbol {k+q}}\,c^+_{\boldsymbol {k'{\mbox{-}}q}}\,c_{\boldsymbol k'}\,c_{\boldsymbol k}\, ,
\end{equation}
where $V_q=4\pi e^2/(V\epsilon_b q^2)$ is simply the 3D Fourier transform of the Coulomb potential. The nature of
the correlation function in Eq. (122) leads to introduce the Green function defined by
\begin{equation}
G({\boldsymbol k}{{\boldsymbol k}'}{\boldsymbol q}, t)=-\,i\,\theta(t)\,
\big <\big [c^+_{\boldsymbol {k{\mbox{-}}q}}(t)\,c_{\boldsymbol k}(t) ,
c^+_{{\boldsymbol k}'}(0)\,c_{\boldsymbol{k'{\mbox{-}}q}}(0) \big ]\big >\, ,
\end{equation}
where $\theta(t)$ is the Heaviside step function. The equation of motion for the Green function is obtained by differentiating both sides of Eq. (124) with respect to time. The result is
\begin{eqnarray}
i\,\frac{\partial}{\partial t}G({\boldsymbol k}{{\boldsymbol k}'}{\boldsymbol q}, t)&=&
\delta(t)\,
\big <\big [c^+_{\boldsymbol {k{\mbox{-}}q}}(t)\,c_{\boldsymbol k}(t) ,
c^+_{{\boldsymbol k}'}(0)\,c_{\boldsymbol{k'{\mbox{-}}q}}(0) \big ]\big > \nonumber\\
&+& \theta(t)\,
\Big <\Big [\frac{\partial}{\partial t} \Big (c^+_{\boldsymbol {k{\mbox{-}}q}}(t)\,c_{\boldsymbol k}(t)\Big ) ,
c^+_{{\boldsymbol k}'}(0)\,c_{\boldsymbol{k'{\mbox{-}}q}}(0) \Big ]\Big > \, .
\end{eqnarray}
We now use the Heisenberg equation of motion [with the operator $\mathcal {O}=
c^+_{\boldsymbol {k{\mbox{-}}q}}(t)\,c_{\boldsymbol k}(t)$]
\begin{equation}
i\hbar\frac{\partial \mathcal {O}}{\partial t}=[\mathcal {O}, H]
\end{equation}
in order to eliminate the time derivative of the operator $\mathcal {O}$ in Eq. (125).  As a result, we obtain
\begin{eqnarray}
i\,\hbar\,\frac{\partial}{\partial t}G({\boldsymbol k}{{\boldsymbol k}'}{\boldsymbol q}, t)&=&
\hbar\,\delta(t)\,
\big <\big [c^+_{\boldsymbol {k{\mbox{-}}q}}(t)\,c_{\boldsymbol k}(t) ,
c^+_{{\boldsymbol k}'}(0)\,c_{\boldsymbol{k'{\mbox{-}}q}}(0) \big ]\big > \nonumber\\
&-& i\, \theta(t)\,
\big <\big [\underbrace{\big [\big (c^+_{\boldsymbol {k{\mbox{-}}q}}(t)\,c_{\boldsymbol k}(t)\big ), H\big ]} ,
c^+_{{\boldsymbol k}'}(0)\,c_{\boldsymbol{k'{\mbox{-}}q}}(0) \big ]\big > \, .
\end{eqnarray}
The (underbraced) commutator in this equation has terms involving products of six fermion operators. Consequently, the
second term on the rhs of this equation is proportional to a higher-order Green function that turns out to contain
products of six fermion operators. One could develop the equation of motion for this higher-order Green function with
the result of an infinite hierarchy of differential equations. It is not usually easy to do infinite-order perturbation
theory on the way, as the (analytical) process becomes extremely cumbersome. This does not, however, mean that the
formalism is useless: one can, in fact, often extract relatively simple closed-form answers by the use of a decoupling
scheme in which, at some stage, the chain of equations is broken off and the higher-order Green functions are expressed
(approximately) in terms of lower-order Green functions. In order to avoid expanding on many lengthy and involved mathematical steps, we wish to put forward here a thoughtful strategy to solve Eq. (127), which straightens the tough
task and saves us time and space. The process to solve Eq. (127) requires us to follow these steps: we (i) need to
take extreme care while opening the underbraced commutator particularly when solving the part that involves the Coulomb interactions and employ EMA whenever needed, (ii) decouple the equations of motion by making use of the RPA, (iii) make rigorous use of the rules of second quantization such as, e.g., $c^+_i\,c_j +c_j\,c^+_i=\delta_{ij}$ and $\big < c^+_{\boldsymbol k}(t)c_{\boldsymbol k'}(t)\big >= f({\boldsymbol k})\,\delta_{{\boldsymbol k}{\boldsymbol k'}}$, where $f({\boldsymbol k})=[e^{\beta[\epsilon_{\boldsymbol k}-\epsilon_F]}+1]^{-1}$ is the Fermi distribution function with
$\beta=(k_B T)^{-1}$ as the inverse temperature and $\epsilon_F$ the Fermi energy, and (iv) exploit further the RPA to
factor the thermal averages, retaining only those terms which force ${\boldsymbol q}={\boldsymbol q'}$. As a result, Eq. (127) simplifies to
\begin{eqnarray}
i\,\hbar\,\frac{\partial}{\partial t}G({\boldsymbol k}{{\boldsymbol k}'}{\boldsymbol q}, t)&=&
\hbar\,\delta(t)\,\delta_{{\boldsymbol k}{\boldsymbol k'}}
[f({\boldsymbol {k{\mbox{-}}q}})-f({\boldsymbol k})]\nonumber\\
&+& [\epsilon_{{\boldsymbol k}}-\epsilon_{{\boldsymbol {k{\mbox{-}}q}}}]\,
G({\boldsymbol k}{{\boldsymbol k}'}{\boldsymbol q}, t)\nonumber\\
&+&V_{\boldsymbol q}\,[f({\boldsymbol {k{\mbox{-}}q}})-f({\boldsymbol k})]\,\sum_{{\boldsymbol {k}_1}}\,
G({\boldsymbol {k}_1}{{\boldsymbol k}'}{\boldsymbol q}, t)\, .
\end{eqnarray}
The solution of this equation is facilitated by introducing the Fourier transform of $G({\boldsymbol {k}_1}{{\boldsymbol k}'}{\boldsymbol q}, t)$ with respect to time:
\begin{equation}
G({\boldsymbol {k}}{{\boldsymbol k}'}{\boldsymbol q}, t)=\frac{1}{2\pi}\,
\int d\omega\, e^{-i\omega t}\,G({\boldsymbol {k}}{{\boldsymbol k}'}{\boldsymbol q}, \omega)\, .
\end{equation}
Taking Fourier transform of both sides of Eq. (128) then yields the integral equation
\begin{equation}
G({\boldsymbol k}{{\boldsymbol k}'}{\boldsymbol q}, \omega)=
\hbar\,\mathbb{S}({\boldsymbol k}\,{\boldsymbol q},\omega)\delta_{{\boldsymbol k}{\boldsymbol k'}}
+ V_q\,\mathbb{S}({\boldsymbol k}\,{\boldsymbol q},\omega)\,\sum_{{\boldsymbol {k}_1}}
G({\boldsymbol {k}_1}{{\boldsymbol k}'}{\boldsymbol q}, \omega)\, ,
\end{equation}
where
\begin{equation}
\mathbb{S}({\boldsymbol k}\,{\boldsymbol q},\omega)=
\frac{f({\boldsymbol {k{\mbox{-}}q}})-f({\boldsymbol k})}
{\epsilon_{{\boldsymbol {k{\mbox{-}}q}}}-\epsilon_{{\boldsymbol k}}+\hbar\omega^+}\, .
\end{equation}
Equation (130) can be solved by the standard procedure of treating an integral equation with a degenerate kernel. The
result is a double-time Green function defined by
\begin{equation}
G({\boldsymbol k}{{\boldsymbol k}'}{\boldsymbol q}, \omega)=
\hbar\,\mathbb{S}({\boldsymbol k}\,{\boldsymbol q},\omega)\delta_{{\boldsymbol k}{\boldsymbol k'}}
+ V_q\,\frac{\mathbb{S}({\boldsymbol k}\,{\boldsymbol q},\omega)
                \mathbb{S}({\boldsymbol {k}'}\,{\boldsymbol q},\omega)}
{1-V_q\, \sum_{{\boldsymbol {k}_1}}\, \mathbb{S}({\boldsymbol {k}_1}{\boldsymbol q}, \omega)}\, .
\end{equation}
Let us redefine the retarded Green function in Eq. (124) formally by
\begin{equation}
G(t)=-i\,\theta(t)\,\big <\big [A(t), B(0) \big ] \big >\, ,
\end{equation}
where $A(t)=c^+_{\boldsymbol {k{\mbox{-}}q}}(t)\,c_{\boldsymbol k}(t)$ and
$B(0)=c^+_{{\boldsymbol k}'}(0)\,c_{\boldsymbol{k'{\mbox{-}}q}}(0)$. Then the scattering cross-section can be safely
related to the Fourier transformed correlation function
\begin{equation}
J(t)=\big <A(t)\,B(0) \big >
\end{equation}
We now know that if the system is in thermal equilibrium, the Fourier transforms of the two quantities are related by
the fluctuation-dissipation theorem [66]
\begin{equation}
J(\omega)=-2\, [n(\omega)+1]\, {\rm Im}[G(\omega)]
\end{equation}
where $n(\omega)=[e^{\beta \hbar \omega}-1]^{-1}$ is the Bose-Einstein distribution function. Substituting Eq. (135)
in Eq. (122) yields
\begin{eqnarray}
\frac{d^2 \sigma}{d\omega d\Omega}
=-\frac{1}{\pi}\, r^2_0\, \Big (\frac{\omega_s}{\omega_i}\Big )\,(\hat {e}_i \centerdot \hat{e}_s)^2\,[n(\omega)+1]\,
{\rm Im}
\Big [\sum_{{\boldsymbol k}{\boldsymbol k}'}\,G({\boldsymbol k}{{\boldsymbol k}'}{\boldsymbol q}, \omega)\Big ] .
\end{eqnarray}
Next, let the single-particle density-density correlation function
\begin{equation}
\chi^0({\boldsymbol q}, \omega)=\frac{1}{V}\,\sum_{{\boldsymbol k}}\,
          \mathbb{S}({\boldsymbol k}{\boldsymbol q}, \omega)
\end{equation}
and the nonlocal, dynamic dielectric function
\begin{equation}
\epsilon({\boldsymbol q}, \omega)= 1- V'_q\,\chi^0({\boldsymbol q}, \omega)\, ,
\end{equation}
where $V'_q=4\pi e^2/(\epsilon_b q^2)$. Then Eq. (132) can be cast in the following form:
\begin{eqnarray}
\sum_{{\boldsymbol k}{\boldsymbol k'}}\,G({\boldsymbol k}{{\boldsymbol k}'}{\boldsymbol q}, \omega)
&=&V\,\hbar\,\Big [\chi^0({\boldsymbol q}, \omega) + V'_q\,
      \frac{[\chi^0({\boldsymbol q}, \omega)]^2}{\epsilon({\boldsymbol q}, \omega)} \Big ]\nonumber\\
&=&V\,\hbar\,\chi^0({\boldsymbol q}, \omega)\,\epsilon^{-1}({\boldsymbol q}, \omega)\nonumber\\
&=&V\,\hbar\,\chi({\boldsymbol q}, \omega)\, .
\end{eqnarray}
Here $\chi ({\boldsymbol q}, \omega)$ is the 3D Fourier transform of the interacting density-density correlation
function $\chi ({\boldsymbol r},{\boldsymbol r'}, \omega)$. Substituting
$\sum_{{\boldsymbol k}{\boldsymbol k'}}\,G({\boldsymbol k}{{\boldsymbol k}'}{\boldsymbol q}, \omega)$ from Eq.
(139) into eq. (136) leaves us with the differential scattering cross-section per unit volume
\begin{eqnarray}
\frac{d^2 \sigma'}{d\omega d\Omega}
&=&-\,\frac{\hbar}{\pi}\, r^2_0\, \Big (\frac{\omega_s}{\omega_i}\Big )\,
      (\hat {e}_i \centerdot \hat{e}_s)^2\,[n(\omega)+1]\,
                   {\rm Im}\big [\chi({\boldsymbol q}, \omega)\big ]\nonumber\\
&=&+\,r^2_0\, \Big (\frac{\omega_s}{\omega_i}\Big )\,
      (\hat {e}_i \centerdot \hat{e}_s)^2\,[n(\omega)+1]\, S({\boldsymbol q}, \omega)\, .
\end{eqnarray}
In writing the second equality, we have exploited the well-known relation between the interacting DDCF
$\chi ({\boldsymbol q}, \omega)$ and the dynamical structure factor (DSF) $S ({\boldsymbol q}, \omega)$:
${\rm Im} \big [ \chi ({\boldsymbol q}, \omega)\big ]
= -(\pi/\hbar)\,\big [S({\boldsymbol q}, \omega) - S({\boldsymbol q}, -\omega)\big ]$.
This is a particularly useful form for computing the differential scattering cross-section. It is observed that the scattering cross-section for the ILS is basically proportional to the imaginary part of the Fourier transform of
the interacting DDCF $\chi(...)$ or to Fourier transform of the DSF $S(...)$. Let us not
forget that in obtaining Eq. (140) we have kept the formulation complying with the
conventional (3D) system. Fortunately, however, the only quantity that needs to be {\em conformed} according to the
system at hand is the DDCF $\chi ({\boldsymbol q}, \omega)$ [see, e.g., Sec. II.G]. Just as in the case of IES, the
prefactors do not matter much because they can only influence the height, width, or (possibly a little) shape of
the Raman intensity, but not the {\em position} in energy. We think that this bare-bone simple case gives a
valuable perspective on the treatment of real semiconducting systems and leaves the option wide open to generalize
the theory to include the influence of, e.g., the anisotropy, an applied magnetic field, the spin-orbit interactions,
the complex band structure, ...etc.

In the nonresonant theories as designed above, for example, which assume the photon to be inetracting entirely with
the conduction electrons, the resonant Raman scattering (RRS) intensity is proportional to the dynamical structure
factor (DSF) pertaining to the conduction electrons and therefore has peaks at the collective (plasmon) excitations
(CPE) at the appropriate wavelengths. Confining to the polarized RRS geometry indicating the absence of spin flips
in the electronic excitations, the DSF peaks should correspond to the poles of the interacting DDCF. Next, the single-particle excitations (SPE), which occur at the poles of the noninteracting DDCF, carry no long-wavelength
spectral weight and hence should not, in principle, show up in the polarized RRS. Astonishingly, the experimental observation, however, is that there always is a relatively weak (but distinguishable) low-energy SPE-like peak in
the RRS spectra, in addition to the expected high-energy CPE peaks. This unexpected existence of the SPE-like peak
[in quasi-n-dimensional systems, with $n\lesssim 2$] has bothered a few authors [67, 68] who chose to include the
role of the valence band electrons by considering the interband transitions and predicted the SPE-like peaks in the
polarized and depolarized geometries for certain laser energies. They argue that the wave-vector dependence of the
said peak intensities is different in the resonant and nonresonant situation explaining only qualitatively the
experimental results. Interestingly enough, the theoretical scheme of Ref. 68 has also garnered a reasonable
experimental support [69]. This briefly sheds light on the so-far-only-known relevance of the interband transitions
on the Raman spectra in the quasi-n-dimensional systems.

\subsection{The analytical diagnoses}

This section is devoted to derive and simplify some necessary analytical results in order to help support the
computation of, for example, the collective excitation spectrum, the density of states, the Fermi energy, the
loss function in the inelastic electron scattering, and the Raman intensity in the inelastic light scattering,
... etc. The most important aspects of the general strategy requires to define three legitimate concerns: the
temperature, the subband occupancy, and the nature of the confining potential. This is systematically discussed
in what follows.

\subsubsection{The zero temperature limit}

It is becoming widely known that electronic, optical, and transport experiments in low-dimensional systems are
performed at extremely low temperatures. Therefore, we choose to confine ourselves to zero temperature limit.
The zero temperature limit has certain advantages: (i) this allows us to replace the Fermi distribution function
with the Heaviside unit step function, i.e., $f(\epsilon)=\theta(\epsilon_F - \epsilon) =1$ (0) for
$\epsilon_F >$ ($<$) $\epsilon$; with $\epsilon_F$ being the Fermi energy in the system, (ii) this leads us to
change the sum over ${\boldsymbol k}$ to an integral by using the summation replacement convention with respect
to 2D such as, e.g., $\sum_{\boldsymbol k}\to \big [A/(2\pi)^2\big ]
\big [\int^{{\boldsymbol k}_F}_0 d{\boldsymbol k}=\int^{k_F}_0 dk\,k \int^{2\pi}_0 d\phi \big ]$, and, most
importantly,  (iii) this allows us to go further and calculate analytically the manageable forms of the
polarizability function $\Pi_{nn'}(...)$. At $T=0$ K, $\Pi_{nn'}(...)$ in Eq. (14) becomes
\begin{eqnarray}
\Pi_{nn'}({\boldsymbol q}, \omega)
=\frac{2}{A}\,\sum_{\boldsymbol k}\,
\Big [\frac{1}{\epsilon_{{\boldsymbol k},n}-\epsilon_{{\boldsymbol {k+q}},n'}+\hbar\omega^+} -
\frac{1}{\epsilon_{{\boldsymbol {k{\mbox{-}}q}},n'}-\epsilon_{{\boldsymbol k},n}+\hbar\omega^+}\Big ]\, ,
\end{eqnarray}
where we have dropped the subscript `$\|$' over ${\boldsymbol k}$ and ${\boldsymbol q}$ for the sake of brevity and
where
\begin{eqnarray}
\epsilon_{{\boldsymbol k},n} - \epsilon_{{\boldsymbol {k+q}},n'}
=-\,\frac{\hbar^2 q^2}{2m^*} - \Delta_{n'n} - \frac{\hbar^2 k q}{m^*}\,\cos(\phi)\nonumber\\
\epsilon_{{\boldsymbol {k{\mbox{-}}q}},n'} - \epsilon_{{\boldsymbol k},n}
=+\,\frac{\hbar^2 q^2}{2m^*} + \Delta_{n'n} - \frac{\hbar^2 k q}{m^*}\,\cos(\phi)\, ,
\end{eqnarray}
where $\phi$ is the angle between ${\boldsymbol k}$ and ${\boldsymbol q}$ and
$\Delta_{n'n}=\epsilon_{n'}-\epsilon_{n}$ is the subband spacing -- with the notion that $n'>n$. Converting the sum
into integral and rearranging terms properly yields
\begin{eqnarray}
\Pi_{nn'}({\boldsymbol q}, \omega)
&=&\frac{m^*k_F}{2\pi^2 \hbar^2 q}\,\int^1_0 dx\, x\,\int^{2\pi}_{0}
         \Big [\frac{1}{u_{-}-x \cos(\phi)} - \frac{1}{u_{+}-x \cos(\phi)}\Big ] \nonumber\\
&=&\frac{m^*k_F}{\pi \hbar^2 q}\,\big[F(u_{-}) - F(u_{+}) \big ]\, ,
\end{eqnarray}
where
\begin{equation}
F(u_{\pm})=u_{\pm} - \sqrt{u_{\pm}-1}
\end{equation}
and
\begin{equation}
u_{\pm}=\frac{\hbar\omega^+ \pm \Delta_{n'n}}{\hbar q v_F} \pm \frac{q}{2k_F}
\end{equation}
Here $k_F$ and $v_F$ are, respectively, the Fermi wave vector and the Fermi velocity in the problem. It is important
to notice that the feasible solutions require that the square root of a complex quantity is always chosen to be the
one with positive imaginary part. In the long wavelength limit (i.e., $q\to 0$),
$\Pi_{nn}\simeq \frac{n_{n}\, q^2}{m^* \omega^2}$ and
$\chi_{nn'} [=\Pi_{nn'}+\Pi_{n'n}]\simeq \frac{2\,(n_{n}-n_{n'})\,\Delta}{(\hbar \omega)^2-\Delta^2}$; $n_n$ being
the 2D electron density in the $n$th subband.  It is interesting to note that these long wavelength limits of the polarizability functions are independent of the dimensionality of the system [1]. We estimate the temperature
dependence of our results would be significant only at $T \gtrsim 35$ K.

\subsubsection{Limiting the number of subbands}

The dielectric function matrix to be generated by Eq. (20) is, in general, an $\infty \times \infty$ matrix until and
unless we restrain the number of subbands (and hence limit the electronic transitions) involved in the problem.
In addition, it is noteworthy that while experiments may report multiple subbands
occupied, theoretically it is extremely difficult to compute the excitation spectrum for the multiple-subband model.
The reason is that the generalized dielectric function turns out to be a matrix of the dimension of
$\eta^2 \times \eta^2$, where $\eta$ is the number of subbands in the model. Handling such enormous matrices (for a
very large $\eta$) analytically is a {\em hard nut to crack} and then no (new, interesting, or important) fundamental
science or technology is ever known to be emerging out of such undue complexity [1]. For this reason, we choose to
keep the complexity to a minimum and limit ourselves to a two-subband model ($n, n', m, m' \equiv 1, 2$) with only the
lowest one occupied. This is quite a reasonable assumption for these low-density, low-dimensional systems at lower temperatures where most of the (electronic, optical, and transport) experiments are carried out. This implies that the generalized dielectric function is to be a $4 \times 4$ matrix
\begin{equation}
\tilde{\epsilon}(q,\omega)=
\begin{bmatrix}
1-\Pi_{11}\,F_{1111} \ & \ -\Pi_{11}\,F_{1112} \ & \ -\Pi_{11}\,F_{1121} \ & \ -\Pi_{11}\,F_{1122} \\
-\Pi_{12}\,F_{1211} \ & \ 1-\Pi_{12}\,F_{1212} \ & \ -\Pi_{12}\,F_{1221} \ & \ -\Pi_{12}\,F_{1222} \\
-\Pi_{21}\,F_{2111} \ & \ -\Pi_{21}\,F_{2112} \ & \ 1-\Pi_{21}\,F_{2121} \ & \ -\Pi_{21}\,F_{2122} \\
-\Pi_{22}\,F_{2211} \ & \ -\Pi_{22}\,F_{2212} \ & \ -\Pi_{22}\,F_{2221} \ & \ 1-\Pi_{22}\,F_{2222}
\end{bmatrix}\, .
\end{equation}
Note that $\Pi_{22}=0$, since the second subband is unoccupied. As the quasi-particle excitations are given by
$\big | \tilde{\epsilon}(q,\omega) \big |=0$, Eq. (146) finally yields
\begin{equation}
(1-\Pi_{11}\,F_{1111})\,(1-\chi_{12}\,F_{1212}) - \Pi_{11}\,\chi_{12}\,F^2_{1112}=0\, ,
\end{equation}
where $\chi_{12}=\Pi_{12}+\Pi_{21}$ is the intersubband polarizability function that takes account of both upward
and downward transitions. This is the final equation to be treated at the computational level in order to obtain
the single-particle as well as collective (plasmon) excitation spectrum in the Q-DEG system at hand. Notice,
however, that further simplification may arise depending upon the nature of the confining potential (see next).

\subsubsection{Symmetry of the confining potential}

It is broadly known that for a symmetric potential well, $F_{ijkl}(q_{\|})$ (the Fourier-transform of the Coulombic
interactions) is strictly zero provided that $i+j+k+l$ is an {\em odd} number [1]. This is so because the
corresponding wave function is either symmetric or antisymmetric under space reflection. Since the 1D harmonic
potential confining the carrier motion along the z direction is symmetric, $F_{1112}(q_{\|})=0$ in Eq. (147). This
implies that the intrasubband and intersubband excitations represented, respectively, by the first and second
factors in the first term in Eq. (147) are decoupled, because the (second) coupling term is zero. Since the subband
index 0 (1) is allowed for the intrasubband (intersubband) excitations, we still need to conform Eq. (147) such that
the subscript $1 \to 0$ and $2 \to 1$ for all practical purposes. The explicit expressions for $F_{0000}(q_{\|})$, $F_{0101}(q_{\|})$, and $F_{0001}(q_{\|})$ as derived from Eq. (17) are given by
\begin{equation}
F_{0000}(q_{\|})=
\frac{2 e^2}{\epsilon_b q_{\|}}\,\int dx\,\int dx'\, e^{-x^2}\,e^{-q_{\|}\ell_c\mid x-x'\mid}\, e^{-x'^{2}}\, ,
\end{equation}
\begin{equation}
F_{0101}(q_{\|})=
\frac{4 e^2}{\epsilon_b q_{\|}}\,\int dx\,\int dx'\, x\,x'\,e^{-x^2}\,e^{-q_{\|}\ell_c\mid x-x'\mid}\,
e^{-x'^{2}}\, ,
\end{equation}
and
\begin{equation}
F_{0001}(q_{\|})=
\frac{2\sqrt{2} e^2}{\epsilon_b q_{\|}}\,\int dx\,\int dx'\,x'\, e^{-x^2}\,e^{-q_{\|}\ell_c\mid x-x'\mid}\,
e^{-x'^{2}}.
\end{equation}
It should be pointed out that the symbols $x=z/\ell_c$ and $x'=z'/\ell_c$ in these equations are dimensionless
variables. We have plotted $F_{0000}(q_{\|})$, $F_{0101}(q_{\|})$, and $F_{0001}(q_{\|})$ as a function of reduced
momentum transfer $q_{\|}/k_F$ in Fig. 5. The results clearly substantiate the notion as stated above and hence
lead us to infer that $F_{0001}(q_{\|})=0$, just as expected. Also noteworthy is the fact that the $F_{0000}(q_{\|})$
remains predominant over the whole range of propagation vector. This is clearly attributed to the charge-carrier
concentration wholly piled up in the lowest subband.

\begin{figure}[htbp]
\includegraphics*[width=8cm,height=9cm]{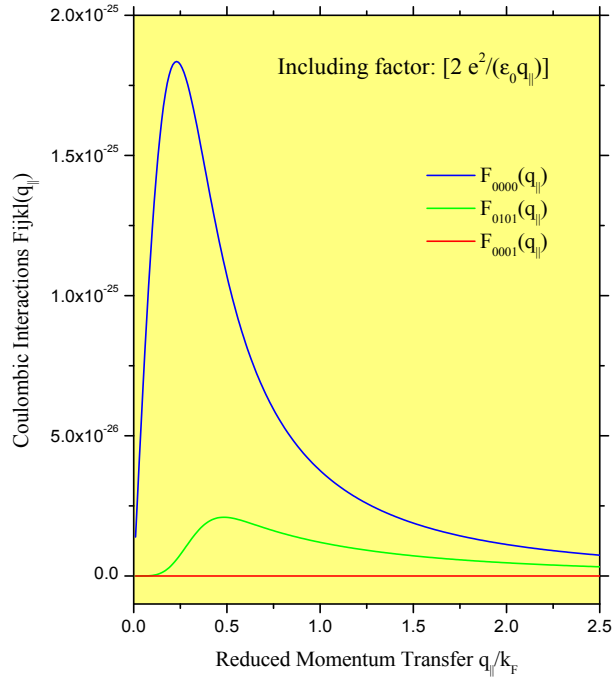}
\caption{The Fourier transformed Coulombic interactions $F_{0000}(q_{\|})$, $F_{0101}(q_{\|})$, and
$F_{0001}(q_{\|})$ plotted as a function of the reduced momentum transfer $q_{\|}/k_F$. We call
attention to the $F_{0001}(q_{\|})$ (in red) which is practically zero over the whole range of
propagation vector.}
\label{fig5}
\end{figure}

\subsubsection{With respect to $\epsilon^{-1}({\boldsymbol q}_{\|}, \omega; z, z')$ in the IES}

Since the whole formulation for the inelastic electron scattering is finally represented in terms of the
${\rm Im}[\epsilon^{-1}(...)]$, it is thought to be important to simplify a few steps which remain quite involved
in Sec. II.C. To do this, we need to exploit our strategy of a limiting two-subband model.
This means that the composite index $\mu, \nu =nn'$ can take only three values 11, 12, and 21. It is needless to
say that $nn'$ cannot take the value 22 because this would lead to a situation that would be self-nullified since
$\Pi_{22}(...)=0$ for the reason that the second subband is unoccupied. Let us recall Eq. (38) and emphasize that
\begin{equation}
{\mbox {the matrix}} \,\,\,\,\, \tilde{\Lambda}= {\mbox {the matrix}} \,\,\,\,\,
          \big [\tilde{I} - \tilde{\Pi}\tilde{\beta} \big ]^{-1}\, .
\end{equation}
This then enables us to diagnose the inverse dielectric function in Eq. (41). A careful analysis leads us to write
\begin{equation}
\sum_{\mu\nu}\,L^*_{\mu}(z)\,\Pi_{\mu}\,\Lambda_{\mu\nu}\,S_{\nu}(z')=
L^*_{11}(z)\,P_{11}\,S_{11}(z') + L^*_{12}(z)\,P_{12}\,S_{12}(z')\, ,
\end{equation}
where
\begin{equation}
P_{11}=\frac{\Pi_{11}}{1-\Pi_{11}\,\beta_{1111}}\, ,
\end{equation}
and
\begin{equation}
P_{12}=\frac{\chi_{12}}{1-\chi_{12}\,\beta_{1212}}\, .
\end{equation}
As stated above, we still need to conform these equations such that the subscript $1 \to 0$ and $2 \to 1$ for any
and every practical purpose. The reader may very well ask why not simply write $P_{11}$ as $P_{00}$ and $P_{12}$
as $P_{01}$ -- just as in the case of $F_{ijkl}(q_{\|})$ above -- and just forget making this bizarre statement.
In part, such a question does make sense. What we, however, implicitly intend to mean by this is that, unlike the
present strategy [of working with a harmonic confining potential which allows the subband index $n$ to take the
values 0 (1) for the ground (first excited) state] there may be the case [where, for instance, the confining
potential is the square-well type] where the subband index {\em has} to take the values 1 (2) for the ground (first
excited) state. Let us leave this elementary doctrine right here: a few things in life can be learnt better from
practice than from preaching. The aforesaid simplification, Eq. (152), turns out to be very useful in carrying out
the computation for the IES phenomena.

\subsubsection{With respect to $\chi ({\boldsymbol q}_{\|}, \omega; q_z)$ in the ILS}

In this section, we intend to search a manageable expression for the interacting DDCF [$\chi ({\boldsymbol q}, \omega)
=\chi ({\boldsymbol q}_{\|}, \omega; q_z)$] involved in Eq. (140) used to compute the (light scattering) differential cross-section or simply the Raman intensity $I(\omega)={\rm Im}[\chi ({\boldsymbol q_{\|}}, \omega; q_z)]$. For this
purpose, we first expand $\chi ({\boldsymbol q_{\|}}, \omega; z, z')$ in terms of the wave functions in the z direction
such that
\begin{equation}
\chi ({\boldsymbol q}_{\|}, \omega; z, z')=\sum_{ijkl}\, \chi_{ijkl}({\boldsymbol q}_{\|}, \omega)\,
\phi_i(z)\,\phi_j(z)\,\phi_k(z')\,\phi_l(z')\, .
\end{equation}
We do not need to worry about the fact that we are no longer using the asterisks appropriately on the confining wave functions and not putting them in the order they should be. The former concern remains immaterial because they all
are real functions [see, e.g., Eq. (3)] and the latter because the spin-orbit interactions have been neglected. It is
not difficult to prove, from the Dyson equation [see Eq. (9)], that
\begin{equation}
\chi_{ijkl}({\boldsymbol q}_{\|}, \omega)=\chi^0_{ij}({\boldsymbol q}_{\|}, \omega)\,\delta_{ik}\,\delta_{jl} + \chi^0_{ij}({\boldsymbol q}_{\|}, \omega)\,\sum_{mn}\,F_{ijmn}({\boldsymbol q}_{\|})\,
\chi_{mnkl}({\boldsymbol q}_{\|}, \omega)\, ,
\end{equation}
where $\chi^0_{ij}({\boldsymbol q}_{\|}, \omega)$ is the matrix element of the DDCF in the absence of the Coulombic
interactions defined by
\begin{equation}
\chi^0 ({\boldsymbol q}_{\|}, \omega; z, z')=\sum_{ij}\, \chi^0_{ij}({\boldsymbol q}_{\|}, \omega)\,
\phi_i(z)\,\phi_j(z)\,\phi_i(z')\,\phi_j(z')\, .
\end{equation}
Next, we define
\begin{eqnarray}
\chi({\boldsymbol q}_{\|}, \omega; q_z)
&=&\int dz\, \int dz'\, e^{-iq_z(z-z')}\,\chi({\boldsymbol q}_{\|}, \omega; z, z')\nonumber\\
&=&\sum_{ijkl}\,\chi_{ijkl}({\boldsymbol q}_{\|}, \omega)\,B_{ijkl}(q_z)\, ,
\end{eqnarray}
with
\begin{equation}
B_{ijkl}(q_z)=\int dz\, \int dz'\,\phi_i(z)\,\phi_j(z)\,\big[e^{-iq_z(z-z')}\big]\,\phi_k(z')\,\phi_l(z')\, .
\end{equation}
Similarly, we define
\begin{eqnarray}
\chi^0({\boldsymbol q}_{\|}, \omega; q_z)
&=&\int dz\, \int dz'\, e^{-iq_z(z-z')}\,\chi^0({\boldsymbol q}_{\|}, \omega; z, z')\nonumber\\
&=&\sum_{ij}\,\chi^0_{ij}({\boldsymbol q}_{\|}, \omega)\,C_{ij}(q_z)\, ,
\end{eqnarray}
with
\begin{equation}
C_{ij}(q_z)=\int dz\, \int dz'\,\phi_i(z)\,\phi_j(z)\,\big[e^{-iq_z(z-z')}\big]\,\phi_i(z')\,\phi_j(z')\, .
\end{equation}
Herein comes an issue that needs to be clarified before we proceed further. The first equalities in Eqs. (158) and
(160) clearly indicate that we are Fourier transforming Eqs. (155) and (157) with respect to the z coordinate,
which represents direction of confinement. This means that there is a lack of translational invariance along the z
direction in the space and hence one should {\em not}, as a matter of principle, seek the Fourier transform of
these quantities. And yet, we violate the fundamental concept [of Fourier transformation] for the benefit of
mathematical convenience. The question, however, is: Do we have a choice? The answer, unfortunately, is {\em no}.
We do not have a choice because the Raman peaks in the ILS experiments are not a function of the spatial positions
in the system; we can only interpret them in terms of the Fourier transforms of the correlation functions. For a
two-subband model, as is the case here, it turns out that $B_{1111}=C_{11}$ and $B_{1212}=C_{12}$. Equation (156),
for a two-subband model, gives rise to a $4\times 4$ matrix whose only nonvanishing elements are 1,1; 2,2; 2,3;
3,2; and 3,3 -- the rest of them vanish for two obvious reasons: (i) due to the second subband being unoccupied,
and (ii) due to the symmetry of the confining potential (see above). It is not so much difficult to determine
$\chi ({\boldsymbol q}_{\|}, \omega; q_z)$ which turns out to acquire a remarkably simple form
\begin{eqnarray}
\chi ({\boldsymbol q}_{\|}, \omega; q_z)
&=&\chi_{1111}\, B_{1111} + \big[\chi_{1212}+\chi_{1221}+\chi_{2112}+\chi_{2121}\big]\,B_{1212}\nonumber\\
&=&\frac{\chi^0_{11}}{1-\chi^0_{11}\,F_{1111}}\,B_{1111} +
   \frac{\chi^0_{12}+\chi^0_{21}}{1-\big (\chi^0_{12}+\chi^0_{21}\big )\,F_{1212}}\,B_{1212}
\end{eqnarray}
This is the final form of $\chi ({\boldsymbol q}, \omega)=\chi ({\boldsymbol q}_{\|}, \omega; q_z)$ to be exploited
in studying the Raman scattering cross-section expressed in Eq. (140). It is interesting to note from Eq. (162)
that (in the two-subband model) the Raman scattering intensity $I(\omega)$ is the sum of two clearly distinguishable
parts: the intrasubband and intersubband collective (plasmon) excitations.

\subsubsection{The Density of states and the Fermi energy}

Independently of the shape and size of a system, the density of states (DOS) is unarguably a defining characteristic
of the behavior of a system and is paramount to the understanding of electronic, optical, and transport phenomena in
the condensed matter physics. The same is true of the Fermi energy because all the transport properties of a system
are a mirror reflection of the electron dynamics at/near the Fermi surface in the system. We start with Eq. (4) to
derive the following expression for computing self-consistently the density of states
\begin{equation}
D(\epsilon)=\frac{m^*}{\pi\,\hbar^2}\,\sum_{n}\,\theta (\epsilon - \epsilon_n)
\end{equation}
and the Fermi energy
\begin{equation}
\sqrt{n_{2D}}=\Big (\frac{m^*}{\pi\,\hbar^2}\Big )^{1/2}\,\sum_{n}\,
     \big (\epsilon_F - \epsilon_n \big )^{1/2}\, \theta (\epsilon_F - \epsilon_n)\, ,
\end{equation}
where $k_F=\sqrt{2\pi n_{2D}}$ is the Fermi wave vector and $n_{2D}$ is the areal density (i.e., the number of
electrons per unit area) of the system. It is generally customary to subtract the zero-point energy $\epsilon_0$
($=\frac{1}{2}\hbar\omega_0$) from the Fermi energy to compute the effective Fermi energy of the system. Figure 6
illustrates the (effective) Fermi energy as a function of the 2D charge density (red) and the DOS as a function of
excitation energy (blue): left (right) axis bears the color of the respective property in the picture. We consider
the most widely exploited GaAs/Ga$_{1-x}$Al$_{x}$As system. It has been known for a long time that the 2D density of
states does not depend on the energy and that it takes on the staircase-like function as is evident from Eq. (163).
However, it does not seem to have been demonstrated before that the Fermi energy shows typical (periodic) dips which
lie exactly midway on the horizontal plateaus of the DOS. Note that the larger the confinement potential energy ($\hbar\omega_0$), the smaller the number of such dips. That is why we choose smaller $\hbar\omega_0$ (=4.5 meV) in
order to demonstrate larger number of dips in the Fermi energy. For $\hbar\omega_0 =9.5$ meV, the Fermi energy
observes only two such dips lying above the center of the corresponding two staircases in the DOS.

\begin{figure}[htbp]
\includegraphics*[width=8cm,height=9cm]{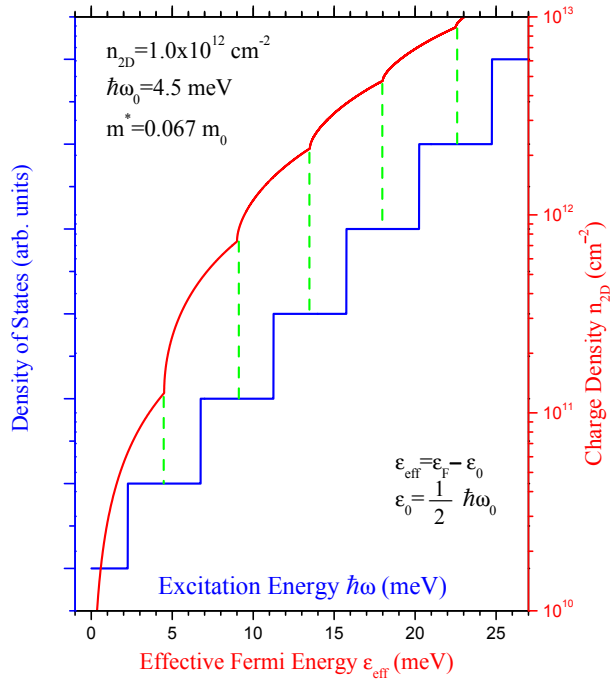}
\caption{(Color Online) The density of states vs. the excitation energy [in Blue] and the effective Fermi energy vs.
the charge density [in red]. We call attention to the respective y-axis denoted by the same color as the DOS or the
Fermi energy.}
\label{fig6}
\end{figure}

\section{Illustrative Numerical Examples}

For the illustrative numerical examples, we focus on the single quantum well in the GaAs/Ga$_{1-x}$Al$_{x}$As system.
The material parameters used are: effective mass $m^*=0.067 m_{_0}$, the background dielectric constant $\epsilon_{_b}
=12.8$, the subband spacing $\hbar\omega_{_0}=9.50$ meV, the self-consistently determined effective Fermi energy $\epsilon_{eff}=9.457$ meV for a 2D charge density $n_{2D}=1.0 \times10^{12}$ cm$^{-2}$, and the effective
confinement width of the parabolic potential well, estimated as the FWHM from the extent of the Hermite function,
$w_{eff}=2\sqrt{2 \ln (2)}\sqrt{n+1}\,\ell_{_c}=21.883$ nm. Notice that the Fermi energy $\epsilon_F$ varies when the
charge density ($n_{2D}$) and/or the confining potential ($\hbar \omega_{0}$) is varied. Thus we aim at discussing
the single-particle and collective excitations, inelastic electron scattering, and inelastic light scattering in a
quasi-2DEG in a two-subband model in the absence of an applied magnetic field within the full RPA at T=0 K. The case
of a nonzero (finite) magnetic field is deferred to a future publication.

\begin{figure}[htbp]
\includegraphics*[width=8cm,height=9cm]{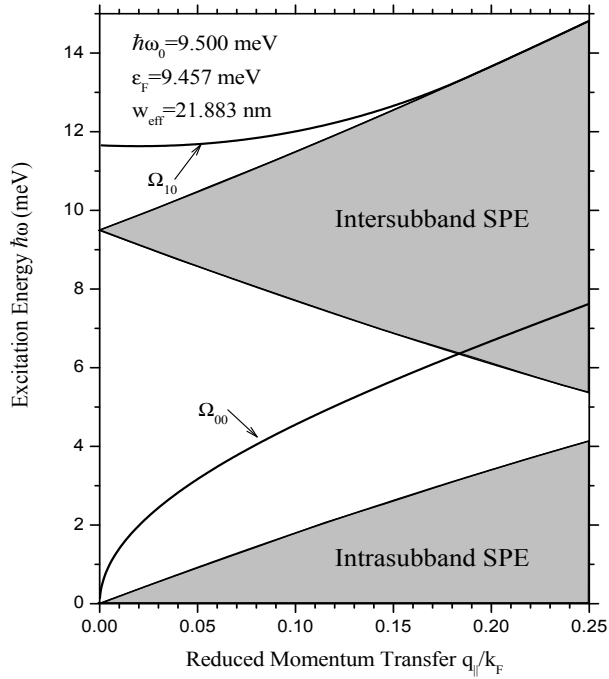}
\caption{The excitation spectrum of a quantum well within a two-subband model where the energy $\hbar \omega$ is
plotted as a function of the reduced momentum transfer $q_{\|}/k_F$. The lower (upper) shaded region refers to
the intrasubband (intersubband) SPE associated with the lowest occupied (first excited) subband at $T=0$ K. The
bold lower (upper) curve marked as $\Omega_{00}$ ($\Omega_{10}$) represents the intrasubband (intersubband) CPE.
Once the CPE fall within the respective SPE, they become Landau-damped and cease to exist as the bona-fide
plasmons with no life-time at all. The relevant parameters are as listed inside the picture.}
\label{fig7}
\end{figure}

\subsection{Excitation spectrum}

Figure 7 illustrates the full excitation spectrum in a quasi-2DEG within a two-subband model in the framework of
Bohm-Pines' RPA. We plot the excitation energy $\hbar\omega$ as a function of dimensionless momentum transfer
$q_{\|}/k_F$. The excitation spectrum is made up of single-particle and collective (plasmon) excitations: the
lower (upper) shaded region stands for the intrasubband (intersubband) single-particle excitations (SPE) -- a
continuum where the polarizability function $\Pi (...)$ and hence the nonlocal, dynamic dielectric function
$\epsilon (...)$ happen to have nonzero imaginary parts. The bold lower (upper) curve marked as $\Omega_{00}$
($\Omega_{10}$) represents the intrasubband (intersubband) collective (plasmon) excitations (CPE). The existence
of the CPE is perturbed inside the respective SPE: when the CPE falls within the SPE continuum, it becomes
Landau-damped and ceases to exist as a bonafide plasmon mode. The intrasubband CPE starts from the origin and does
not seem to merge anywhere with the corresponding SPE. As such, this CPE remains free from Landau damping and is
thus a long-lived, bonafide plasmon excitation. The intersubband CPE starts at ($q_{\|}/k_F=0$,
$\hbar\omega=11.652$ meV), propagates to observe a minimum at ($q_{\|}/k_F=0.02$, $\hbar\omega=11.632$ meV), and
merges with the upper edge of the intersubband SPE at ($q_{\|}/k_F=0.187$, $\hbar\omega=13.369$ meV) to become
Landau-damped thereafter. While it is too much to expect the analytical estimate of the critical point(s) of the
{\em collective} excitations, it is not difficult to check analytically why the intersubband {\em single-particle}
excitation starts at the subband spacing [i.e. at ($q_{\|}/k_F=0$, $\hbar\omega=9.50$ meV)].

It is important to mention that even though a fair portion of the intrasubband plasmon propagates -- after ($q_{\|}/k_F=0.183$, $\hbar\omega=6.376$ meV) -- within the intersubband SPE, the former does {\em not} bear the
brunt of the latter. This is not because the intrasubband and intersubband excitations are decoupled in this
particular case (owing to the {\em symmetry} of the confining potential), rather, in fact, because the two
excitations have different lineages that do not interbreed with one another. Another important issue here is the
energy shift of the intersubband CPE from the respective SPE: this difference is crucially attributed to the
many-body effects such as depolarization and excitonic shifts [1, 15], which are weak enough in the quantum wells
(as compared to those in the quantum wires and the quantum dots).

\subsection{Inelastic electron scattering}

In this section, we compute and discuss the loss functions $P (q_{\|}, \omega)$ derived in Eqs. (83), (90), and (96),
respectively, for the parallel, perpendicular, and shooting-through configurations. It should be underscored that in
all the illustrative examples presented here we have ignored the prefactors [outside the signs of respective
integrals] because, as stated above, their inclusion can only bring an insignificant feature to the loss peaks in the electron energy-loss spectrum.

\begin{figure}[htbp]
\includegraphics*[width=8cm,height=9cm]{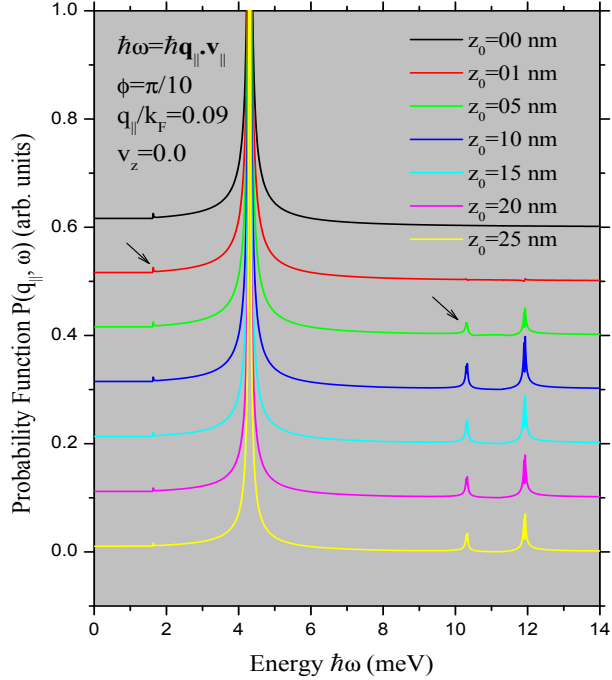}
\caption{(Color Online) The computed loss function $P (q_{\|}, \omega)$ vs. the energy $\hbar\omega$ for a fast-particle
moving parallel to the Q-2DEG in the GaAs/Ga$_{1-x}$Al$_{x}$As quantum well. Each curve corresponds to a different value
of $z_0$ -- the distance between the center of coordinate system and the electron beam. The parameters listed in the
picture are: the reduced momentum transfer $q_{_{\|}}/k_F=0.09$, the perpendicular component of the fast-particle
velocity $v_z=0$, and the angle $\phi$ [defined by ${\bf q}_{_{\|}}\cdot {\bf v}_{_{\|}}=q_{_{\|}}\,v_{_{\|}}\,cos(\phi)$]
specified as $\phi=\pi/10$. Each curve is displaced vertically for the sake of clarity. Note that the y axis is scaled
with no loss of generality. The rest of the parameters are the same as in Fig. 7.}
\label{fig8}
\end{figure}

\subsubsection{The parallel configuration}

Figure 8 illustrates the loss function $P (q_{\|}, \omega)$ as a function of the energy $\hbar\omega$ for a fast-particle moving in the parallel configuration for a Q-2DEG in the GaAs/Ga$_{1-x}$Al$_{x}$As quantum well in the inelastic electron scattering. Notice that this case is somewhat special in the sense that the energy loss in the scattering process turns
out to be automatically specified as $\hbar\omega=\hbar {\bf q}_{_{\|}}\cdot {\bf v}_{_{\|}}$. The important features observed from Fig. 8 are the following. The sharp $\delta$-like peaks at $\hbar\omega=4.3064$ meV and at
$\hbar\omega=11.932$ meV substantiate, respectively, the intrasubband plasmon at $\hbar\omega=4.3039$ meV and the intersubband plasmon at $\hbar\omega=11.918$ meV -- for the corresponding value of the momentum transfer -- in Fig. 7;
and we consider this to be an excellent agreement. Let us now have a careful look at the smaller peaks indicated by
arrows. Such lowest peak at $\hbar\omega=1.6351$ meV corresponds closely to the edge of the intrasubband single-particle
continuum occurring at $\hbar\omega=1.626$ meV [see Fig. 7], whereas the third lowest peak at $\hbar\omega=10.313$ meV
lies inside the intersubband single-particle continuum [just below the upper edge at $\hbar\omega=11.281$ meV (see Fig.
7)]. The particle velocity corresponding to these loss peaks [counting from the origin] is found to be
$v_{_{\|}}=1.01 v_F$, $2.06 v_F$, $6.38 v_F$, and $7.37 v_F$, respectively. One can immediately notice that the larger
the distance $z_0$, the smaller the amplitude of the loss function $P (q_{\|}, \omega)$. It is, however, important to
observe that the positions (in energy) of the loss peaks remain independent of the distance $z_0$. The other observations made after extensive computation for a wide range of parameters are: (i) the larger the distance $z_0$, the smaller the
rate of the energy loss ($W'$), just as it is expected intuitively, and (ii) only the fast-particle velocities greater
than the Fermi velocity make sense. To conclude with, we find that the dominant contribution to the loss peaks comes from
the collective (plasmon) excitations.

\begin{figure}[htbp]
\includegraphics*[width=8cm,height=9cm]{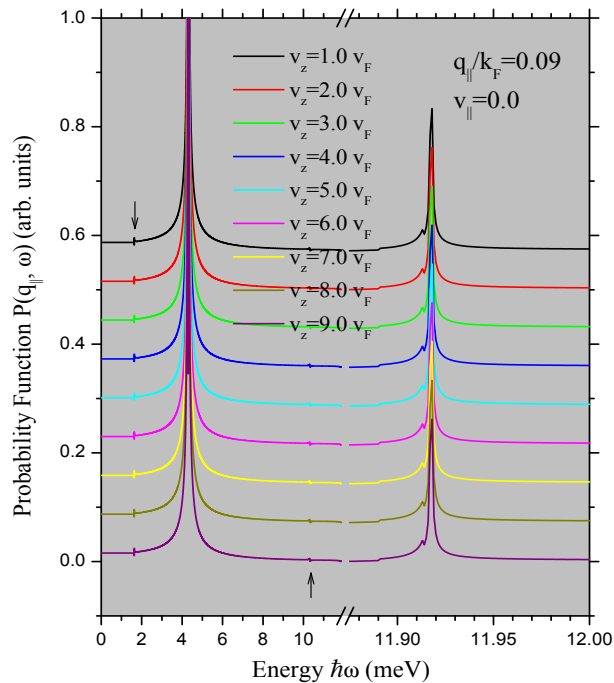}
\caption{(Color Online) The computed loss function $P (q_{\|}, \omega)$ vs. the energy $\hbar\omega$ for a fast-particle
incident at and specularly reflected [with $\theta_i=0=\theta_r$] from the Q-2DEG in the GaAs/Ga$_{1-x}$Al$_{x}$As
quantum well. Each curve corresponds to a different value of particle velocity $v_z$, which changes the sign at time
$t=0$. The parameters listed in the picture are: the reduced momentum transfer $q_{_{\|}}/k_F=0.09$ and the
perpendicular component of the fast-particle velocity $v_{_{\|}}=0$. Each curve is displaced vertically for the sake of clarity. The y axis is scaled with no loss of generality. The scale break of the x axis is noteworthy. The rest
of the parameters are the same as in Fig. 7.}
\label{fig9}
\end{figure}

\subsubsection{The perpendicular configuration}

Figure 9 shows the loss function $P (q_{\|}, \omega)$ as a function of the energy $\hbar\omega$ for a fast-particle
impinging at and specularly reflected [with $\theta_i=0=\theta_r$ (see Fig. 3)] from the surface of a Q-2DEG in the GaAs/Ga$_{1-x}$Al$_{x}$As quantum well in the inelastic electron scattering. The sharp $\delta$-like peaks at $\hbar\omega=4.3045$ meV and at $\hbar\omega=11.918$ meV corroborate, respectively, the intrasubband plasmon at $\hbar\omega=4.3039$ meV and the intersubband plasmon at $\hbar\omega=11.918$ meV, for the corresponding value of the
momentum transfer $q_{_{\|}}/k_F=0.09$ in Fig. 7. Thus there is an outstanding agreement between the loss spectrum
in Fig. 9 and the excitation spectrum in Fig. 7. The (barely visible) weak signals of smaller peaks indicated by arrows
at $\hbar\omega=1.636$ meV and at $\hbar\omega=10.306$ meV tell the similar tale as the corresponding peaks in Fig. 8.
It is worthwhile to note that, for any set of parameters, the total number of loss peaks cannot exceed five, as is
expected in a two-subband model for the Q-2DEG. This remark is valid for any/all configurations subject to a
two-subband model. Again, it is interesting to observe that the $\delta$-like loss peaks (in energy) do not vary with
the variation in the particle velocity. This complies with the fact that the momentum transfer $q_{_{\|}}$ is kept
constant for all the particle velocities.  The previous remark regarding the collective excitations as the primary loss
mechanism still remains valid. A comparative look at Figs. 8 and 9 reveals that the intersubband plasmons become better
observable in the perpendicular geometry.

\begin{figure}[htbp]
\includegraphics*[width=8cm,height=9cm]{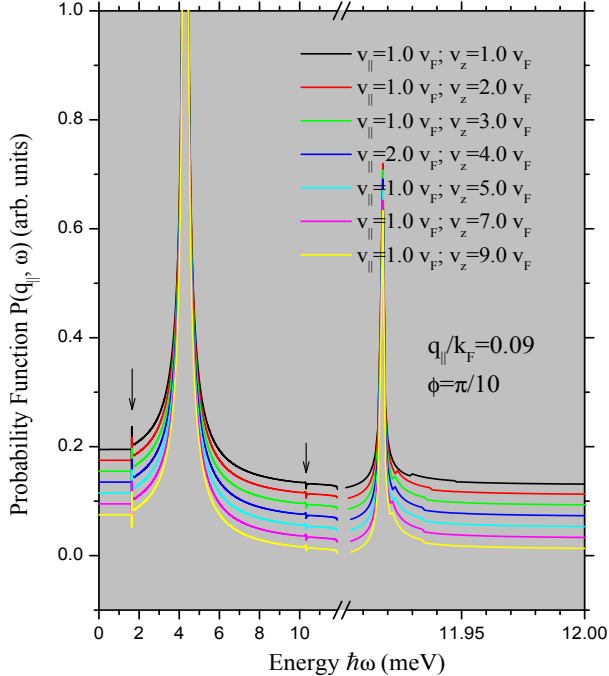}
\caption{(Color Online) The computed loss function $P (q_{\|}, \omega)$ vs. the energy $\hbar\omega$ for a fast-particle
shooting through a Q-2DEG in the GaAs/Ga$_{1-x}$Al$_{x}$As quantum well. Each curve corresponds to a different set of
particle velocity ${\bf v}$ [$=(v_{_{\|}}, v_z)$]. The parameters listed in the picture are: the reduced
momentum transfer $q_{_{\|}}/k_F=0.09$ and the angle $\phi$
specified as $\phi=\pi/10$. Each curve is
displaced vertically for the sake of clarity. The y axis is scaled with no loss of generality. Note the scale
break on the x axis. The rest of the parameters are the same as in Fig. 7.}
\label{fig10}
\end{figure}

\subsubsection{The shooting-through configuration}

Figure 10 depicts the loss function $P (q_{\|}, \omega)$ as a function of the energy $\hbar\omega$ for a fast-particle
shooting through a Q-2DEG in the GaAs/Ga$_{1-x}$Al$_{x}$As quantum well in the inelastic electron scattering. It is
important to notice that in this case we treat the particle velocity ${\bf v}=$ constant ($\Rightarrow$ particle
shoots through). The (comparatively broader but still) sharp $\delta$-like peaks at $\hbar\omega=4.305$ meV and at $\hbar\omega=11.918$ meV validate, respectively, the intrasubband plasmon at $\hbar\omega=4.3039$ meV and the
intersubband plasmon at $\hbar\omega=11.918$ meV, for the corresponding value of the momentum transfer
$q_{_{\|}}/k_F=0.09$ in Fig. 7. This implies an extraordinarily good agreement between the loss spectrum in Fig. 10
and the excitation spectrum in Fig. 7. In addition, the (hardly visible) razor-sharp but weak signals of smaller peaks indicated by arrows at $\hbar\omega=1.636$ meV and at $\hbar\omega=10.304$ meV beat the same drum as the corresponding
peaks in Fig. 8. Again, the positions of the sharp loss peaks pertaining to the intrasubband and intersubband plasmons
remain intact, even though the fast-particle velocity varies. Just like in the previous two geometries, we stress that
the dominant contribution to the loss peaks comes from the collective (plasmon) excitations. An interesting feature
common to both the perpendicular and shooting-through geometries is that, away from the sharp resonances, the loss
function decreases with increasing fast-particle velocity. The rest of the discussion related to previous geometries
is still valid. Notwithstanding that the main physics regarding the loss mechanism is consistent in all three
geometries considered, we feel that this geometry of fast-particle shooting through the Q-2DEG yields, in general,
more pronounced structure in the loss peaks.

Before we close this section, a word is in order regarding the shooting-through geometry. While it is beyond doubt that
the parallel and perpendicular geometries are perfectly in the reach of the current technology, one might feel a little
skeptical about the shooting-through geometry. What may cause such a skepticism is this question: How can one shoot a fast-particle (i.e., a coherent electron beam) through a Q-2DEG embedded in the host material? It is quite likely
that the substrate materials cladding the Q-2DEG would obstruct the fast-particle from passing through the whole system
and reach the detector. Nevertheless, it seems to be a commonsense belief that a highly energetic fast-particle should,
in principle, surmount any such obstacle in its path and shoot through the whole system. Until and unless that day comes,
the shooting-through geometry will remain, at least, of fundamental importance.

\subsection{Inelastic light scattering}

In this section, we discuss the computed Raman intensity $I(\omega)={\rm Im}[\chi ({\boldsymbol q}_{_{\|}}, \omega;
q_z)]$ as a function of excitation energy $\hbar \omega$, for numerous values of the parallel momentum transfer
${\boldsymbol q}_{_{\|}}/k_F$ and for a given value of the normal component $q_z=0.18 k_F$. The calculation of
$I(\omega)$ is equivalent to that of the light scattering cross-section without the prefactors in Eq. (140). Notice
that the computation of $I(\omega)$ provides the full response of the system in the inelastic light scattering from
the electronic excitations in the Q-2DEG as is the case here.

\begin{figure}[htbp]
\includegraphics*[width=8cm,height=9cm]{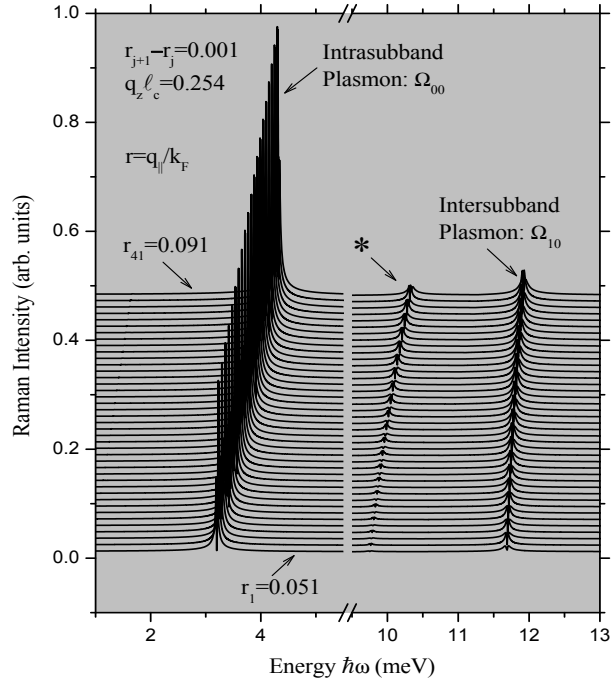}
\caption{The computed Raman intensity $I (\omega)$ vs. the energy $\hbar\omega$ for the inelastic light scattering
from a Q-2DEG in the GaAs/Ga$_{1-x}$Al$_{x}$As quantum well. The long wavelength regime covers
$0.051 \le r=q_{_{\|}}/k_F \le 0.091$ and we feed $q_z \ell_c=0.254\Rightarrow q_z=0.18\, k_F$. The y axis is scaled
with no loss of generality. Each curve is displaced vertically for the sake of clarity. Note the scale break on the
energy (x) axis. The rest of the parameters are the same as in Fig. 7.}
\label{fig11}
\end{figure}

Figure 11 illustrates the Raman intensity $I (\omega)$ as a function of excitation energy $\hbar\omega$ for several
values of the (parallel) momentum transfer in the long wavelength limit specified by
$0.51 \le {\boldsymbol q}_{_{\|}}/k_F \le 0.091$ and for a given value of the (normal) component of the momentum
transfer $q_z=0.18\,k_F $. It is observed that there are three prominent peaks below $\hbar\omega=13$ meV for a
given ${\boldsymbol q}_{_{\|}}/k_F$. The lowest and the third lowest Raman peaks marked as $\Omega_{00}$ and
$\Omega_{10}$ substantiate very clearly the intrasubband and intersubband collective (plasmon) excitations at the corresponding values of ${\boldsymbol q}_{_{\|}}/k_F$, whereas the middle peak (marked by $\ast$) in the Raman
spectrum lies inside the intersubband single-particle continuum just below the upper edge [see, e.g., Fig. 7]. This
leads us to infer that there is an excellent agreement between the Raman spectrum in Fig. 11 and the excitation
spectrum in Fig. 7 regarding the collective excitations. However, this cannot be said about the single-particle
Raman peaks observed in Fig. 11. In addition, we do not see (at least on this scale) any single-particle Raman
peaks that may be comparable to the corresponding intrasubband single-particle peak in the excitation spectrum in
Fig. 7. The complexity regarding the SPE peaks between the RRS experiments and the theoretical excitation spectrum
[such as Fig. 7] is an old puzzle which dates back (almost) four decades and is shared by all systems [see below].


\begin{figure}[htbp]
\includegraphics*[width=8cm,height=9cm]{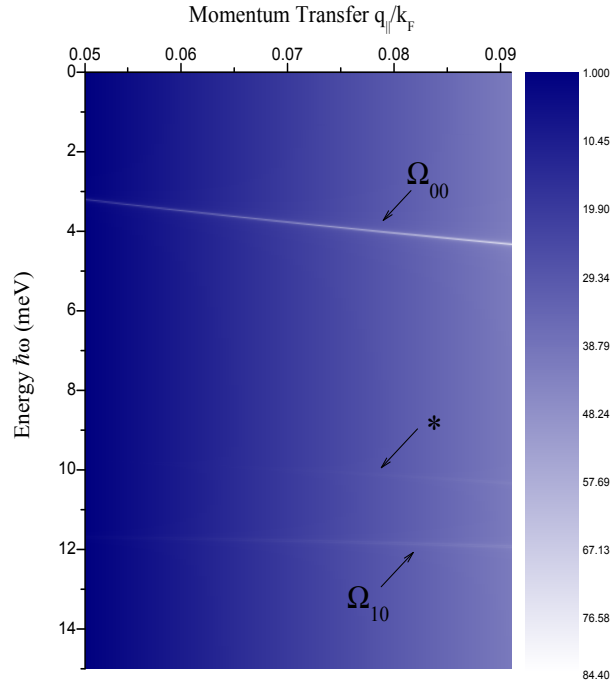}
\caption{The computed Raman intensity $I (\omega)$ vs. the energy $\hbar\omega$ vs. the momentum transfer
$q_{_{\|}}/k_F$  for the inelastic light scattering from a Q-2DEG in the GaAs/Ga$_{1-x}$Al$_{x}$As quantum
well. This 3D figure is far more sure and secure than its 2D counterpart. $q_z=0.18\, k_F$. See Fig. 11 for
other parameters.}
\label{fig12}
\end{figure}

In order to make sure about this perplexing issue of the existence of SPE peaks in the Raman spectrum, we made a
3D plot of the Raman intensity $I (\omega)$, vs. the energy $\hbar\omega$, vs. the (finest mesh of) momentum
transfer ${\boldsymbol q}_{_{\|}}/k_F$. The results are plotted in Fig. 12. This is a more confident way to make
sure if there does (or does not) exist all the expected peaks in the Raman spectrum. What we observe are the
collective (plasmon) excitation peaks (indicated by $\Omega_{00}$ and $\Omega_{10}$) -- both for intrasubband
and intersubband excitations -- and a single-particle peak (indicated by $\ast$) which is found to lie just below
the upper edge inside the intersubband SPE continuum [in Fig. 7]. However, we fail to observe any other SPE peak
in the Raman spectrum shown in Fig. 12.

The issue of the existence of the SPE peaks in the RRS experiments and the intent to explain them was addressed
indirectly in relation with the importance of the interband transitions in Sec. II.F. Here we would like to
touch briefly the issue from a different perspective. The (theoretical) excitation spectrum (TES) (see Fig. 7)
is a result of exploiting the {\em intersubband} spectroscopy entirely within the {\em conduction} band. This
makes clear that TES has absolutely nothing to do with the {\em interband} and/or {\em intervalence} transitions.
The Raman signals in the RRS experiments, on the other hand, are too weak to be detected without making the laser
resonate with the interband transitions (and hence the name RRS). In other words, RRS experiments inherit the
Raman signals from the interband transitions. Therefore, as regards the SPE, the nonresonant theories and the RRS
experiments have, reciprocally, nothing in common. Embodying interband transitions in the Raman intensity
$I (\omega)$ may provide a qualitative explanation [70] of the SPE peaks observed in the RRS experiments. The
experimental verification of single-particle excitations in the nonresonant theories remains a mystery, however.

\section{Concluding Remarks}

In summary, we have investigated thoroughly the electron dynamics of the quasi-2DEG in a quantum well within a
two-subband model in the framework of Bohm-Pines' full RPA. Starting with the single-particle eigenfunction
and eigenenergy for the confining harmonic potential, we provide a systematic route to formulate the nonlocal,
dynamic dielectric function, the inverse dielectric function, the nonlocal screened potential, and the Dyson
equation for studying the interacting DDCF. This provided us with the base to develop in a consistent manner
the theory of the inelastic electron scattering (IES) and of the inelastic light scattering (ILS) in the
quantum wells. Sec. II.G on the analytical diagnoses is an (unusual) bonus adding grace to the whole process of
curiosity. The strength of this work lies more on how and why and less on what.

The illustrative examples embark on the consequence of the symmetry of the confining potential, the variation
of the density of states and the Fermi energy, the presentation of the full excitation spectrum covering both
intrasubband and intersubband -- single-particle as well as collective (plasmon) -- excitations with a
discussion of the Landau damping, the loss spectra for the three principal [parallel, perpendicular, and
shooting-through] geometries in the inelastic electron scattering, and the Raman spectrum depicting the Raman
intensity vs. the (laser) energy in the inelastic light scattering. However, we have not intended to generate
and expand on the illustrative (numerical) results and hence kept to the aim of providing a thorough, yet
concise, methodological tools to think through a research problem.

It is quite interesting to mention that the derivation of the inverse dielectric function (IDF) is important and
its usefulness is multifold. Since the zeros of the dielectric function and the poles of the IDF yield exactly
identical results, the latter not only provides an interesting alternative of the former for studying, for
example, the elementary electronic excitations, it also serves an exclusively important purpose for exploring
the inelastic electron scattering (as we have already seen). The IDF is useful for investigating not only the
electronic and optical phenomena but also the transport ones. For instance, the imaginary (real) part of the IDF
sets to furnish a significant measure of the longitudinal (Hall) resistance in the system. Last but not least,
the exact IDF as derived here knows no bounds with respect to the subband occupancy.

General features worth adding to the problem are: the effect of (i) an applied magnetic field, (ii) the spin-orbit interactions, (iii) the many-body (such as exchange-correlation) effects, and (iv) the coupling to the optical
phonons, to mention a few. Even with the feeling of lack and limitation of the RRS experiments [see, e.g., Sec.
III.C], it should be worthwhile to include the interband transitions in the general expression of the Raman
scattering cross-section, Eq. (140). We feel enthusiastic enough to close by stating that the electron energy-loss
spectroscopy can be a potential alternative of the overused optical spectroscopy such as RRS for investigating the
elementary electronic excitations in the low-dimensional nanostructures.

Finally, the old conviction that theory should not challenge what is yet to be explored experimentally is no longer
a doctrine in practice. Science does not seem to have grown following such beliefs. In the development of science,
there are times when theory lags behind experiment and there are times when the converse is true. What matters is
the verifiability of facts. Currently, we have been extending this whole formalism to be applicable to the quantum
wires and the results will be reported shortly.


\newpage


\begin{acknowledgments}
The author feels enormously grateful to Naomi Halas and Peter Nordlander who never tell someone their dreams
are impossible. He is thankful to Christian Sch\"uller for a series of extensive and stimulating communications.
He has enjoyed very interesting communications with H. Ibach, R.F. Egerton, and M.S. Moreno. He is also
indebted to Federico Garcia-Moliner for partially introducing him to the subject of EELS during his sabbatical
year (1993-1994) at CSIC, Madrid, Spain. Words fail him to appreciate Professor F. Barry Dunning's constant
support and encouragement. Finally, he would like to thank Kevin Singh for his unfailing and timely help with
the software during the course of this investigation.
\end{acknowledgments}





\begin{references}
\bibitem[1]{1} For an extensive review of electronic, optical, and transport phenomena in the systems
               of reduced dimensions, such as quantum wells, wires, dots, and  electrically/magnetically
               modulated 2D systems, see M.S. Kushwaha, Surf. Sci. Rep. {\bf 41}, 1 (2001).
\bibitem[2]{2} A.B. Fowler, F.F. Fang, W.E. Howard, P.J. Stiles, Phys. Rev. Lett. {\bf 16}, 901 (1966).
\bibitem[3]{3} F. Stern, Phys. Rev. Lett. {\bf 18}, 546 (1967).
\bibitem[4]{4} A.V. Chaplik, Sov. Phys. JETP {\bf 33}, 997 (1971).
\bibitem[5]{5} P.B. Vischer, L.M. Falicov, Phys. Rev. B {\bf 3}, 2541 (1971).
\bibitem[6]{6} D. Grecu, Phys. Rev. B {\bf 8}, 1978 (1973).
\bibitem[7]{7} K.W. Chiu, J.J. Quinn, Phys. Rev. B {\bf 9}, 4724 (1974).
\bibitem[8]{8} A.L. Fetter, Ann. Phys. NY {\bf 88}, 1 (1974).
\bibitem[9]{9} M. Apostol, Z. Phys. B {\bf 22}, 13 (1975).
\bibitem[10]{10} D.E. Beck, P. Kumar, Phys. Rev. B {\bf 13}, 2859 (1976).
\bibitem[11]{11} A.K. Rajgopal, Phys. Rev. B {\bf 15}, 4262 (1977).
\bibitem[12]{12} D.A. Dahl, L.J. Sham, Phys. Rev. B {\bf 16}, 651 (1977).
\bibitem[13]{13} S. Mori, T. Ando, Phys. Rev. B {\bf 19}, 6433 (1979).
\bibitem[14]{14} N. Tzoar, P.M. Platzman, Phys. Rev. B {\bf 20}, 4189 (1979).
\bibitem[15]{15} T. Ando, A.B. Fowler, F. Stern, Rev. Mod. Phys. {\bf 54}, 437 (1982).
\bibitem[16]{16} D. Olego, A. Pinczuk, A.C. Gossard, andW.Wiegmann, Phys. Rev. B {\bf 25}, 7867 (1982).
\bibitem[17]{17} R. Sooryakumar, A. Pinczuk, A. Gossard, and W. Wiegmann, Phys. Rev. B {\bf 31}, 2578
                 (1985).
\bibitem[18]{18} G. Fasol, N. Mestres, H. P. Hughes, A. Fischer, and K. Ploog, Phys. Rev. Lett. {\bf 56},
                 2517 (1986).
\bibitem[19]{19} A. Pinczuk, M. G. Lamont, and A. C. Gossard, Phys. Rev. Lett. {\bf 56}, 2092 (1986).
\bibitem[20]{20} A. Pinczuk and J. P. Valladares, D. Heiman, A. C. Gossard, J. H. English, C. W. Tu,
                 L. Pfeiffer, and K. West, Phys. Rev. Lett. {\bf 61}, 2701 (1988).
\bibitem[21]{21} A. Pinczuk, S. Schmitt-Rink, G. Danan, J. P. Valladares, L. N. Pfeiffer, and K. W. West,
                 Phys. Rev. Lett. {\bf 63}, 1633 (1989).
\bibitem[22]{22} G. Danan, A. Pinczuk, J. P. Valladares, L. N. Pfeiffer, K. W. West, and C. W. Tu,
                 Phys. Rev. B {\bf 39}, 5512 (1989).
\bibitem[23]{23} D. Gammon, B. U. Shanabrook, J. C. Ryan, and D. S. Katzer, Phys. Rev. B {\bf 41}, 12311
                 (1990).
\bibitem[24]{24} D. Gammon, B. V. Shanabrook, J. C. Ryan, D. S. Katzer, and M. J. Yang, Phys. Rev. Lett.
                 {\bf 68}, 1884 (1992).
\bibitem[25]{25} A. Pinczuk, B. S. Dennis, L. N. Pfeiffer, and K. West, Phys. Rev. Lett. {\bf 70}, 3983
                 (1993).
\bibitem[26]{26} R. Decca, A. Pinczuk, S. Das Sarma, S. Dennis, L. N. Pfeiffer, and K. W. West, Phys.
                 Rev. Lett. {\bf 72}, 1506 (1994).
\bibitem[27]{27} J. Wagner, J. Schmitz, F. Fuchs, J. D. Ralston, P. Koidl, and D. Richards, Phys. Rev. B
                 {\bf 51}, 9786 (1995).
\bibitem[28]{28} D. S. Kainth, D. Richards, A. S. Bhatti, H. P. Hughes, M. Y. Simmons, E. H. Linfield, and
                 D. A. Ritchie, Phys. Rev. B {\bf 59}, 2095 (1999).
\bibitem[29]{29} C. Lohe, A. Leuther, A. Förster, and H. Lüth, Phys. Rev. B {\bf 47}, 3819 (1993).
\bibitem[30]{30} G.R. Bell, C.F. McConville, and T.S. Jones, Phys. Rev. B {\bf 56}, 15995 (1997).
\bibitem[31]{31} E. Fermi, Phys. Rev. {\bf 57}, 485 (1940).
\bibitem[32]{32} H. A. Kramers, Physica {\bf 13}, 401 (1947).
\bibitem[33]{33} R.H. Ritchie, Phys. Rev. {\bf 106}, 874 (1957).
\bibitem[34]{34} R. H. Ritchie and A. L. Marusak, Surf. Sci. {\bf 4}, 234 (1966).
\bibitem[35]{35} A. A. Lucas and M. Sunjic Prog. Surf. Sci. {\bf 2}, 75 (1972).
\bibitem[36]{36} D. L. Mills, Surf. Sci. {\bf 48}, 59 (1975).
\bibitem[37]{37} W.L. Schaich, Phys. Rev. B {\bf 24}, 686 (1981).
\bibitem[38]{38} B.N.J. Persson and E. Zaremba, Phys. Rev. B {\bf 31}, 1863 (1985).
\bibitem[39]{39} R.E. Camley, D.L. Mills, Phys. Rev. B {\bf 29}, 1695 (1984).
\bibitem[40]{40} P. Hawrylak, J.W. Wu, J.J. Quinn, Phys. Rev. B {\bf 32}, 4272 (1985).
\bibitem[41]{41} S.R. Streight and D.L. Mills, Rev. B {\bf 35}, 6337 (1987).
\bibitem[42]{42} G. Gumbs, Phys. Rev. B {\bf 39}, 5185 (1989).
\bibitem[43]{43} V. Z. Kresin and H. Morawitz, Phys. Rev. B {\bf 43}, 2691 (1991).
\bibitem[44]{44} W.H. Backes, F.M. Peeters, F. Brosens, and J.T. Devreese, Phys. Rev. B {\bf 45}, 8437
                 (1992).
\bibitem[45]{45} R.F. Egerton, {\it Electron Energy-Loss Spectroscopy} (Plenum, New York, 1996).
\bibitem[46]{46} P.M. Platzman, Phys. Rev. {\bf 139}, A379 (1965).
\bibitem[47]{47} Y. Yafet, Phys. Rev. {\bf 152}, 858 (1966).
\bibitem[48]{48} P.A. Wolf, Phys. Rev. {\bf 171}, 436 (1968).
\bibitem[49]{49} S.S. Jha, IL Nuovo Cimento {\bf LXIII B}, 331 (1969).
\bibitem[50]{50} D. Hamilton and A.L. McWhorter, in: {\it Light Scattering Spectra of Solids},
                 Ed. G.B. Wright (Springer, New York, 1969).
\bibitem[51]{51} F.A. Blum, Phys. Rev. {\bf 1}, 1125 (1970).
\bibitem[52]{52} M.V. Klein, in: {\it Light Scattering in Solids}, Ed. M. Cardona (Springer,
                 Berlin, 1975).
\bibitem[53]{53} E. Burstein, A. Pinczuk, and D.L. Mills, Surf. Sci. {\bf 98}, 451 (1980).
\bibitem[54]{54} S. Katayama and T. Ando, J. Phys. Soc. Jpn. {\bf 54}, 1615 (1985).
\bibitem[55]{55} J.K. Jain and P.B. Allen, Phys. Rev. B {\bf 32}, 997 (1985).
\bibitem[56]{56} J.K. Jain, S. Das Sarma, Phys. Rev. B {\bf 35}, 918 (1987).
\bibitem[57]{57} M. Cardona and I. P. Ipatova, in: {\it Elementary Excitations in Solids}, Eds.
                 J. L. Birman, C. S\'ebenne, and R. F. Wallis (Elsevier, New York, 1992).
\bibitem[58]{58} E. G. Mishchenko, Phys. Rev. B {\bf 53}, 2083 (1996).
\bibitem[59]{59} L. A. Falkovsky, Phys. Rev. B {\bf 70}, 054301 (2004).
\bibitem[60]{60} C. Schuller, {\it Inelastic Light Scattering of Semiconductor Nanostructures}
                 (Springer, Berlin, 2006).
\bibitem[61]{61} H. Ibach, J. Electron. Spectros. Relat. Phenom. {\bf 64/65}, 819 (1993).
\bibitem[62]{62} D. Pines, {\it The Many-Body Problem} (Benjamin, New York, 1961);
                 A.L. Fetter and J.D. Walecka, {\it Quantum Theory of Many-Particle Systems}
                 (McGraw-Hill, New York, 1971); G.D. Mahan, {\it Many Particle Physics}
                 (Plenum, New York, 1981).
\bibitem[63]{63} M.S. Kushwaha and F. Garcia-Moliner, Phys. Lett. A {\bf 205}, 217 (1995).
\bibitem[64]{64} W. Heitler, {\it Quantum Theory of Radiation} (Oxford University Press, Oxford, 1954).
\bibitem[65]{65} J.M. Luttinger and W. Kohn, Phys. Rev. {\bf 97}, 869 (1955).
\bibitem[66]{66} D.N. Zubarev, Sov. Phys. -- Uspekhi {\bf 3}, 320 (1960).
\bibitem[67]{67} C. Steinebach, C. Schuller, and D. Heitmann, Phys. Rev. B {\bf 59}, 10240 (1999).
\bibitem[68]{68} S. Das Sarma and D.W. Wang, Phys. Rev. Lett. {\bf 83}, 816 (1999).
\bibitem[69]{69} B. Jusserand, M. N. Vijayaraghavan, F. Laruelle, A. Cavanna, and B. Etienne, Phys. Rev.
                 Lett. {\bf 85}, 5400 (2000).
\bibitem[70]{70} C. Sch\"uller, Private Communication.
\end{references}
\end{document}